\documentclass[12pt]{article}
\usepackage{amsfonts}
\usepackage{amssymb}
\usepackage{graphicx}
\usepackage{setspace}
\usepackage{natbib}
\usepackage{amsmath}
\usepackage{amsthm}
\usepackage{palatino}
\usepackage{paralist}
\usepackage{subcaption}
\usepackage{xcolor}
\usepackage{multirow}
\usepackage{booktabs}
\usepackage{dsfont}
\usepackage{indentfirst}
\usepackage{enumerate}
\usepackage{longtable}
\usepackage{caption,bbm}
\usepackage{algpseudocode} 
\usepackage{tikz}
\usepackage{lscape}
\usepackage{graphicx}
\usepackage{caption}
\usetikzlibrary{arrows}
\usetikzlibrary{positioning}
\usetikzlibrary{calc}
\usepackage{adjustbox}
\usetikzlibrary{positioning,arrows.meta}
\usepackage[ruled,linesnumbered]{algorithm2e}
\usepackage{algorithm2e,algpseudocode}
\newdimen\nodeDist
\defcitealias{han2024tensor}{Han et al. (2024a)}
\defcitealias{han2024cp}{Han et al. (2024b)}

\usepackage[hyperindex,breaklinks]{hyperref}
\hypersetup{colorlinks=true,       
	linkcolor=red,       
	citecolor=blue,        
	filecolor=magenta,      
	urlcolor=cyan           
}  

\usepackage{multirow}
\usepackage[toc,page]{appendix}
\usepackage{setspace}
\usepackage{threeparttable}

\usepackage[left = 1in, right = 1in, top = 1in, bottom = 1in]{geometry}

\AtBeginDocument{
  \setlength{\abovedisplayskip}{3pt}
  \setlength{\belowdisplayskip}{3pt}
  \setlength{\abovedisplayshortskip}{3pt}
  \setlength{\belowdisplayshortskip}{3pt}
}
\usepackage{titlesec}
\titleformat*{\section}{\large\bfseries}
\titleformat*{\subsection}{\normalsize\bfseries}
\titleformat*{\subsubsection}{\normalsize\bfseries}
\titleformat*{\paragraph}{\normalsize\bfseries}

\titlespacing{\section}{0pt}{*1.5}{*1.5}
\titlespacing{\subsection}{0pt}{*1.5}{*1.5}
\titlespacing{\subsubsection}{0pt}{*1.2}{*1.2}
\titlespacing{\paragraph}{0pt}{*1.5}{*1.5}

\newtheorem{theorem}{Theorem}

\newtheorem{corollary}{Corollary}
\newtheorem{remark}{Remark}
\newtheorem{assumption}{Assumption}

\newcommand{\utwi}[1]{\mbox{\boldmath $ #1$}}

\newcommand{\bb}{{\utwi{b}}}

\newcommand{\bA}{{\utwi{A}}}
\newcommand{\bB}{{\utwi{B}}}

\newcommand{\bE}{{\utwi{E}}}
\newcommand{\bF}{{\utwi{F}}}

\newcommand{\bI}{{\utwi{I}}}

\newcommand{\bM}{{\utwi{M}}}

\newcommand{\bP}{{\utwi{P}}}
\newcommand{\bQ}{{\utwi{Q}}}

\newcommand{\bS}{{\utwi{S}}}

\newcommand{\bU}{{\utwi{U}}}
\newcommand{\bV}{{\utwi{V}}}
\newcommand{\bW}{{\utwi{W}}}

\newcommand{\bY}{{\utwi{Y}}}
\newcommand{\bZ}{{\utwi{Z}}}

\newcommand{\hbU}{\widehat{\utwi{U}}}

\newcommand{\hbZ}{\widehat{\utwi{Z}}}
\newcommand{\hbM}{\widehat{\utwi{M}}}

\newcommand{\hbC}{\widehat{\utwi{C}}}
\newcommand{\hbB}{\widehat{\utwi{B}}}
\newcommand{\hbS}{\widehat{\utwi{S}}}

\newcommand{\hbW}{\widehat{\utwi{W}}}

\newcommand{\bLambda}{{\utwi{\mathnormal\Lambda}}}

\newcommand{\bbeta}{{\utwi{\mathnormal\beta}}}

\newcommand{\bfeta}{{\utwi{\mathnormal\eta}}}

\newcommand{\cE}{{\cal E}}
\newcommand{\cF}{{\cal F}}

\newcommand{\cS}{{\cal S}}

\newcommand{\cX}{{\cal X}}

\def\mathbold{\boldsymbol}

\def\E{\mathbb{E}}
\def\P{\mathbb{P}}
\def\RR{\mathbb{R}}
\def\R{\mathbb{R}}
\def\mat{\hbox{\rm mat}}

\title{\bf \Large{Panel Coupled Matrix-Tensor Clustering Model with Applications to Asset Pricing}\thanks{\scriptsize{
We thank Biao Cai, Lilun Du, and seminar and conference participants at City University of Hong Kong
for invaluable comments and discussions. 
Cui (E-mail: \texttt{liyuan.cui@cityu.edu.hk}), Feng (E-mail: \texttt{gavin.feng@cityu.edu.hk}), and Li (E-mail: \texttt{jiayali6-c@my.cityu.edu.hk}) are at the City University of Hong Kong; Han (E-mail: \texttt{yuefeng.han@nd.edu} is at the University of Notre Dame. Han is the corresponding author. Co-advisors are ordered alphabetically.
}}
}

\author{\normalsize{Liyuan Cui} \and \normalsize{Guanhao Feng} \and \normalsize{Yuefeng Han} \and \normalsize{Jiayan Li}}

\date{\small \today}

\begin{document}
	
\maketitle
\vspace{-1cm}

\begin{abstract}
\onehalfspacing \noindent 
We tackle the challenge of estimating grouping structures and factor loadings in asset pricing models, where traditional regressions struggle due to sparse data and high noise. 
Existing approaches, such as those using fused penalties and multi-task learning, often enforce coefficient homogeneity across cross-sectional units, reducing flexibility.
Clustering methods (e.g., spectral clustering, Lloyd's algorithm) achieve consistent recovery under specific conditions but typically rely on a single data source.
To address these limitations, we introduce the Panel Coupled Matrix-Tensor Clustering (PMTC) model, which simultaneously leverages a characteristics tensor and a return matrix to identify latent asset groups. By integrating these data sources, we develop computationally efficient tensor clustering algorithms that enhance both clustering accuracy and factor loading estimation.
Simulations demonstrate that our methods outperform single-source alternatives in clustering accuracy and coefficient estimation, particularly under moderate signal-to-noise conditions.
Empirical application to U.S. equities demonstrates the practical value of PMTC, yielding higher out-of-sample total $R^2$ and economically interpretable variation in factor exposures.

\medskip

	
\end{abstract}

\newpage

\section{Introduction}
Common factor models are central to empirical asset pricing, capturing time-series co-movement and cross-sectional return variation \citep[e.g.,][]{fama_five-factor_2015}.
However, estimating asset-specific betas (factor loadings) remains statistically challenging; high idiosyncratic volatility in sparse datasets often masks underlying risk signals, a problem exacerbated in segmented markets.
Market segmentation leads to significant cross-sectional variation in factor risk premia \citep{hou2011factors}, limiting the explanatory power of a common factor model across asset classes.
To address this, \citet{patton2022risk} and \citet{cong2023sparse} propose different clustering methods to group assets by within-group factor loadings, revealing significant cross-sectional heterogeneity. Similarly, \citet{GIGLIO2024TestAssetsandWeakFactors} show that a factor's explanatory power depends on test asset selection, as factor loadings vary across asset classes. These results highlight the potential of group-specific factor models to better explain variations in asset returns.
However, these group-specific factor models often rely solely on asset returns ($Y_{t}$) and factor returns ($f_t$), neglecting asset characteristics ($\cX_{ t}$), which provide valuable incremental information \citep{kelly2019IPCA}.

From a statistical perspective, these group-specific models can be viewed as a matrix clustering task, modeling excess returns of $p_1$ assets using $m_1$ factors with latent groups:
\begin{equation}\label{mode:y}
\bY = \bM_1 \bB \bF + \boldsymbol{\eta},
\end{equation}
where $\bY \in \mathbb{R}^{p_1 \times T}$ is the return matrix, $\bF \in \mathbb{R}^{m_1 \times T}$ are factors, $\bB \in \mathbb{R}^{r_1 \times m_1}$ denotes group-level loadings, and $\bM_1 \in \{0,1\}^{p_1 \times r_1}$ encodes asset memberships. 
Clustering methods, including $k$-means and spectral clustering \citep[e.g.,][]{jain2010data, vonluxburg2007tutorial, zhang2024leave}, and extensions to structured/high-order data \citep[e.g., stochastic block models and tensor clustering, ][]{gao2022iterative, han2022exact}, provide consistent recovery under suitable conditions but typically rely on a single data source.

Asset pricing researchers observe returns $\bY$, factors $\bF$, and asset-specific characteristics that contain information beyond returns alone \citep[e.g., see the tensor application in ][]{lettau20243d}. 
These characteristics naturally form a tensor with latent group structure:
\begin{equation}\label{mode:x}
\mathcal{X} = \mathcal{S} \times_1 \bM_1 \times_2 \bM_2 + \mathcal{E}.
\end{equation}
where {$\mathcal{X}\in \mathbb{R}^{p_1\times p_2\times T}$ collects $p_2$ asset characteristics for $p_1$ stocks over $T$ periods}, 
$\bM_2 \in \mathbb{R}^{p_2 \times r_2}$ is the membership matrix for characteristics, and $\cS \in \mathbb{R}^{r_1 \times r_2 \times T}$ is a core tensor capturing cluster centroids. The shared first mode, $\bM_1$, provides a direct link between the outcome matrix $\bY$ and the characteristics tensor $\cX$. Here the $k$-mode product of $\cX\in\RR^{p_1\times p_2\times \cdots \times p_K}$ with a matrix $\bU\in\RR^{r_k\times d_k}$, denoted as $\cX\times_k \bU$, is an order $K$-tensor of size $d_1\times \cdots \times d_{k-1} \times r_k\times d_{k+1}\times \cdots \times d_K$ such that
$ (\cX\times_k \bU)_{i_1,...,i_{k-1},j,i_{k+1},...,i_K}=\sum_{i_k=1}^{d_k} \cX_{i_1,i_2,...,i_K} \bU_{j,i_k}.$

Asset characteristics are typically incorporated in two approaches. 
The first follows the classical portfolio‐sorting scheme \citep[e.g.,][]{Fama1992crosssection}, where a small set of characteristics, size and value, for example, is used to sort individual assets into portfolios and evaluate factor models under the implicit assumption that all assets within a portfolio share the same loading vector $\bbeta$. This approach, however, relies on only a limited subset of available characteristics, depends on ad hoc sorting breakpoints, and may introduce selection bias. The second approach uses conditional factor models, which specify factor loadings as explicit functions of characteristics, as in \citet{kelly2019IPCA} and \citet{gu2021autoencoder}. While flexible, these conditional models use characteristics to define loadings rather than to identify latent asset groupings.
While asset characteristics are known to proxy for risk exposures and expected returns, traditional sorting and conditional models fail to fully integrate this high-dimensional information into a unified grouping structure.
Our paper proposes a complete integration of the return matrix and asset characteristic tensor, enhancing the accuracy of cross-sectional clustering and factor loading estimation compared to methods that rely on a single return source.

In this paper, we propose a coupled matrix-tensor clustering framework 
integrates the heterogeneous factor model \eqref{mode:y} with a characteristics tensor \eqref{mode:x} by enforcing a shared membership matrix ($\bM_1$) across both data sources.
Our objective is to jointly exploit the shared clustering structure of the matrix and tensor, thereby improving accuracy over single-source approaches.
To achieve this, we develop a two‐stage estimation strategy. 
The first stage employs Panel Coupled Matrix-Tensor Spectral Clustering (PMTSC) to obtain a warm initialization, while the second stage refines the clustering via the Panel Coupled Matrix-Tensor Lloyd (PMTLloyd) algorithm. 
PMTSC relies on a coupled low‐rank factorization, implemented through Panel Coupled High‐Order Orthogonal Iteration (PCHOOI), which extends HOOI to settings where a tensor and a matrix are jointly modeled. PMTLloyd then iteratively updates cluster assignments using innovative orthogonal projection-based refinements. Importantly, even when applied to tensor data alone, PMTLloyd improves upon existing refinement procedures \citep{han2022exact}, indicating that the algorithmic contribution is not limited to the coupled setting. The recovered clustering structure further enables accurate estimation of the group-level factor loading matrix.

Our theoretical analysis reveals that coupling strengthens the signal in the shared mode by increasing relevant matricized singular values, which relaxes signal-to-noise ratio (SNR) requirements under appropriate conditions. 
A particularly striking result is that PMTLloyd achieves sharp misclustering error bounds that are uniformly superior to those of single-source clustering, regardless of the relative signal strengths in the tensor and matrix components. 
Additionally, our error bounds feature exact constants in the exponent, significantly improving upon previous tensor co-clustering results \citep{han2022exact}.
For factor loading estimation, we establish convergence rates under both observed and latent factor scenarios, demonstrating that increases in either $p_1$ or $T$ enhance the estimation of the loading matrix. These bounds dominate ungrouped factor analysis results, with the notable property of guaranteeing consistency even with finite sample sizes, improving over existing group panel regression results \citep{su2016identifying}. Simulation studies confirm that PMTLloyd achieves the lowest clustering error, even in weak SNR regimes, compared with other popular methods.

Empirical results further highlight the practical value of the proposed algorithms when applied to the Panel Tree portfolios of \citet{cong2025growing} for the period 1980-2024. Our method yields a higher out-of-sample total $R^2$ than return-based clustering and traditional univariate or bivariate characteristic sorting methods. By jointly exploiting information in returns and asset characteristics, the method identifies sharper and more stable latent asset groups, resulting in more accurate factor-loading estimates and improved predictive performance. The clusters reveal economically meaningful patterns: differences in factor exposures align with underlying characteristics, providing interpretable links between the empirical ``factor zoo'' and characteristic-driven asset behavior. These findings demonstrate that coupling $\bY$ and $\cX$ enhances clustering precision, providing a more robust and interpretable framework for understanding cross-sectional return variation.

\subsection{Related Literature}

Our paper contributes to recent advances in low-rank tensor decomposition. Tucker-type models offer efficient representations for multi-way data, enabling theoretical analysis of estimation and recovery \citep[e.g.,][]{zhang2018tensor,zhang2019optimal}. Recent progress in tensor clustering has delivered sharp guarantees for identifying structured groups in high-order data under various noise regimes \citep[e.g.,][]{Luo2021Sharp, Hu2022Multiway, Luo2022Tensor,lyu2023optimal, Lyu2025Optimal}. We extend this line of research by introducing a coupled matrix-tensor framework that achieves uniformly superior misclustering error bounds and improved computational efficiency compared to state-of-the-art single-source tensor clustering methods.

In addition to the matrix clustering paradigm, model~\eqref{mode:y} can be viewed from the perspective of coefficient homogeneity. Under this view, assets correspond to separate regression tasks, and the grouping structure induces equality constraints on regression coefficients within each group. This connects our framework to the literature on homogeneity pursuit, where pairwise (fused) penalties recover latent group structures among coefficients \citep[e.g.,][]{shen2010grouping, zhu2013simultaneous}. Recent work extends these ideas to multi-task and panel settings with adaptive and robust formulations \citep[e.g.,][]{duan2023adaptive, cui2025do}.

Our framework also draws on data fusion methodologies that incorporate auxiliary information to improve statistical efficiency, including Covariate-Assisted Sparse Tensor Completion \citep{ibriga2023covariate} and Covariate-Assisted Spectral Clustering \citep{binkiewicz2017covariate}. These approaches demonstrate how covariates or side information can enhance clustering and representation learning in multi-way settings. Building on this insight, our method treats the characteristics tensor as auxiliary linked data, developing a unified estimation framework that jointly exploits shared clustering structure across outcomes and characteristics. By explicitly modeling the shared latent group structure across the outcome matrix and characteristics tensor, our Panel Coupled Matrix-Tensor Clustering (PMTC) model provides a statistically efficient mechanism for information fusion, thereby enhancing recovery of the shared mode.

We also build on coupled factorization methods that jointly model multiple structured datasets. Matrix-matrix approaches \citep[e.g.,][]{lock2013joint,fan2019distributed,tang2021integrated,ma2024optimal}, matrix-tensor factorizations \citep[e.g.,][]{acar2011all,de2017coupled}, and tensor-tensor formulations \citep[e.g.,][]{liu2023joint,chen2025distributed} provide flexible tools for capturing shared latent structures across heterogeneous sources. 
These methodologies illustrate the benefits of coupling information across multiple modes, a principle underlying the proposed PMTC approach.

\subsection{Notation, Preliminaries and Organization}
\label{sec:notation}
Let $[n]$ denote the set $\{1, 2, \ldots, n\}$. For a vector $x = (x_1, \ldots, x_p)^{\top}$, we define its $\ell_q$-norm as $\| x \|_q = (\sum_{i=1}^p |x_i|^q)^{1/q} $ for $ q \geq 1 $. For a matrix $\bA = (a_{i,j}) \in \mathbb{R}^{m \times n}$, denote its singular values as $\lambda_1(\bA) \geq \lambda_2(\bA) \geq...\geq \lambda_{\min\{m,n\}}(\bA) \geq 0 $. The subspace spanned by the first $r$ left singular vectors is denoted as $ \bU_r = {\rm LSVD}_r (\bA) $, and the spectral norm is $\|\bA\|_2 = \lambda_1(\bA)$. We denote the $i$-th row and $j$-th column of $\bA$ as $\bA_{i:} $ and $\bA_{:j} $, respectively. We also use $a \wedge b = \min\{a, b\}$ and $a \vee b = \max\{a, b\}$.

For any two orthonormal matrices $\bU, \hbU \in \mathbb{O}^{m \times r}$, the distance between their column spaces is measured by the spectral norm of their projection difference
$\ell_2(\bU, \hbU) = \|\hbU \hbU^\top - \bU \bU^\top\|_2 = \sqrt{1 - \lambda_r(\bU^\top \hbU)^2},$
which equals the sine of the largest principal angle between the subspaces. Let $\mathrm{vec}(\cdot)$ denote the vectorization operator. The mode-$k$ unfolding (matricization) of tensor $\mathcal{A}$ is defined as $\mat_k(\mathcal{A})$, mapping $\mathcal{A}$ to a matrix in $\mathbb{R}^{m_k \times m_{-k}}$ where $m_{-k} = \prod_{j \neq k}^K m_j$. For example, if $\mathcal{A} \in \mathbb{R}^{m_1 \times m_2 \times m_3}$, then
$(\mat_1(\mathcal{A}))_{i,(j+m_2(k-1))} = (\mat_2(\mathcal{A}))_{j,(k+m_3(i-1))} = (\mat_3(\mathcal{A}))_{k,(i+m_1(j-1))} = \mathcal{A}_{ijk}.$
For a $d$-way tensor $\mathcal{A}$, we define its minimal matricized singular value as $\lambda_{\min}(\mathcal{A}) = \lambda_{\min}(\mat_i(\mathcal{A})),i = 1,...,d,$ the smallest singular value across all mode-$i$ matricizations.

The remainder of the paper is organized as follows. Section \ref{sec:model} develops the PMTC methodology and presents two estimation algorithms: PMTSC and PMTLloyd. Section \ref{sec:theoretical} establishes theoretical properties and convergence guarantees for the proposed estimators. Section \ref{sec:simulation} reports simulation evidence on clustering and factor loading estimation performance. Section \ref{sec:application} applies the method to empirical asset-pricing data. Section \ref{sec:discussion} concludes with a discussion of potential extensions.

\section{Panel Coupled Matrix-Tensor Clustering} \label{sec:model}

\subsection{The Model}

We consider a general Panel Coupled Matrix-Tensor Clustering (PMTC) model for the asset pricing problem. One observes a $(d+1)$-order characteristics tensor $\mathcal{X} \in \mathbb{R}^{p_1 \times \cdots \times p_d \times T}$ and a panel outcome matrix $\mathbf{Y} \in \mathbb{R}^{p_1 \times T}$.
The model takes the form
\begin{align}
\mathcal{X} &= \mathcal{S} \times_1 \bM_1 \times_2 \cdots \times_d \bM_d + \mathcal{E}, \quad
\bY = \bM_1 \bB \bF + \bfeta,
\label{model1}
\end{align}
or equivalent, 
$\cX_t = \cS_t \times_1 \bM_1 \times_2 \cdots \times_d \bM_d + \bE_t, \quad
Y_{i,t} =\bbeta_{i}^\top f_t + \eta_{it}, \quad \bbeta_{i} =  \sum_{k=1}^{r_1} \bb_{k} \cdot \mathbf{1}\{i \in \mathcal{G}_{1k}\}$, 
where $\cS \in \mathbb{R}^{r_1 \times \cdots \times r_d \times T}$ is a low rank core tensor capturing latent block centroids, $\bB \in \mathbb{R}^{r_1 \times m_1}$ is a group-level factor loading matrix, $\bF \in \mathbb{R}^{m_1 \times T}$ contains $m_1$ observed or latent common factor processes, $\mathcal{G}_{1k}$ is the $k$-th group in the mode-1, and $\bb_i $ is the $i$-th row of $\bB$. The error terms are denoted by $\cE$ and $\bfeta$. Each $\bM_i \in \{0,1\}^{p_i \times r_i}$ is a membership matrix that maps the $p_i$ objects in mode $i$ into $r_i$ latent clusters such that $(\bM_i)_{j,a} = \mathbb{I}\{ j \text{-th fiber in mode-}i \text{ belongs to cluster } a\} $, and $r_i \ll p_i$. By construction, every row of $\bM_i$ contains exactly one nonzero entry, indicating the unique cluster assignment of each entity in mode $i$. In the characteristics tensor $\cX$, the temporal mode (mode $d+1$) typically lacks cluster structure, though our framework can accommodate such extensions.
We further define $\bS_Y = \bB \bF$ to represent the centroids matrix of $\bY$. In this formula, our model not only nests the common factor structure but is also sufficiently general to accommodate other cases where $\bY$ is not strictly driven by factors but still admits a latent group representation.

In the existing literature \citep[e.g.][]{patton2022risk, han2022exact}, the group structure of panel units in $\bY$ is typically derived through one of two approaches: either by employing group factor models or by clustering based on the characteristics tensor $\mathcal{X}$. Our framework integrates these two approaches into a unified model \eqref{model1}. By leveraging the group dependency between panel outcomes $\bY$ and characteristics $\cX$, we enable bidirectional information sharing that significantly improves group estimation along the first tensor mode. Furthermore, this proposed coupled dependency framework provides more interpretable results in real applications compared to methods that analyze panel outcomes $\bY$ or characteristics $\cX$ in isolation.

For applications in empirical asset pricing, the main objectives are twofold: (i) to recover accurate estimates of the membership matrices $\bM_1,\bM_2,\cdots, \bM_d$, thereby identifying the latent grouping structures in both $\mathcal{X}$ and $\bY$; (ii) to leverage these estimated groupings for improved estimate on factor loadings $\bB$. 

\subsection{Methodology}
\label{sec:method}
We estimate membership matrices $\bM_i,i=1,...,d$ (clustering) and factor loadings $\bB$ via a two-stage strategy.
The proposed clustering procedure includes two steps: an initialization step using the Panel Coupled Matrix-Tensor Spectral Clustering (PMTSC) procedure, presented in Algorithm \ref{PCSCalg}, and an iterative refinement step using the Panel Coupled Matrix-Tensor Lloyd algorithm (PMTLloyd), presented in Algorithm \ref{PCLloydalg}. 
Following the estimation of group memberships, we employ PCA or the least squares method to estimate the group-level factor loading matrices.

Under PMTC model \eqref{model1}, we aim to jointly estimate the membership matrices $\bM_1, \bM_2, \ldots, \bM_d$, by solving the following optimization problem: 
\begin{align} \label{loss}
(\hbM_i,i=1,...,d)=\min_{{\mathbold M_i}\in\{0,1\}^{p_i\times r_i},i=1,\cdots,d} \left\| \mathcal{X} - \mathcal{S} \times_{i=1}^d \bM_i  \right\|_F^2 + \left\| \bY - \bM_1 \bS_Y \right\|_F^2.
\end{align}
While problem \eqref{loss} is nonconvex and computationally intractable when optimizing all parameters simultaneously, the objective function is convex in each individual parameter when the others are held fixed. This multi-convex structure naturally lends itself to the use of an efficient alternating optimization algorithm, combined with a warm initialization procedure.

A key advantage of problem \eqref{loss} is its ability to exploit the shared clustering structure in the first mode by integrating information from both $\cX$ and $\bY$.
Specifically, for the first mode, we construct the augmented matrix $\bZ_1 = [\mat_1(\cX), \bY] = \bM_1 [\mat_1(\cS)(\otimes_{j=2}^d \bM_j \otimes \bI_T), \bS_Y]$. This formulation demonstrates that $\bM_1$ can be consistently estimated through the joint analysis of $\cX$ and $\bY$, rather than relying on either source independently. 
For modes $i \neq 1$, we define $ \bZ_i = \mat_i(\cX) = \bM_i [\mat_i(\cS)(\otimes_{j=1,j\neq i}^d \bM_j \otimes \bI_T)] $. 
Although estimation of $\bM_i$ for $i \neq 1$ depends on $\cX$, the coupling $\cX$ and $\bY$ improves the accuracy of $\bM_1$.
This improvement propagates to the other tensor modes, resulting in more reliable cluster recovery for the entire model.
This propagation effect underscores a primary advantage of the coupled framework: enhanced recovery in the shared mode strengthens recovery across all remaining modes in practice.

\begin{remark}[Weighted squared loss]
Instead of \eqref{loss}, we can consider a weighted empirical squared loss that balances the contributions of $\cX$ and $\bY$:
\begin{align}\label{loss2}
(\hbM_i,i=1,...,d)=\min_{{\mathbold M_i}\in\{0,1\}^{p_i\times r_i},i=1,\cdots,d} \omega\left\| \cX - \mathcal{S} \times_{i=1}^d \bM_i \right\|_F^2 + \left\| \bY - \bM_1 \bS_Y \right\|_F^2,     
\end{align}
where $\omega$ weights the tensor $\cX$'s contribution. 
This parameter effectively rescales the data sources, allowing the researcher to balance the clustering signal from returns against the signal from characteristics.
The limiting cases are instructive: $\omega \to 0$ reduces to clustering based solely on $\bY$, while $\omega \to \infty$ yields clustering driven entirely by $\cX$. Given this flexibility, our methodological and theoretical development focuses on the baseline case $\omega=1$.
In practice, $\omega$ can be selected through model evaluation criteria such as in-sample or out-of-sample total $R^2$. This weighting strategy proves particularly valuable when $\cX$ and $\bY$ exhibit substantial differences in scale or variability, preventing the noisier component from dominating the clustering objective. 
\end{remark}

\subsubsection{Panel coupled matrix-tensor spectral clustering algorithm}

We first develop a warm initialization procedure that generalizes high-order spectral clustering with specific coupled-mode innovations.
Similar to tensor block models, the core tensor $\cS$ in model \eqref{model1} may exhibit degenerate ranks, meaning the rank of the unfolded matrix $\mat_i(\cS)$ is strictly less than the dimension $r_{i}$.
This characteristic allows us to reformulate model \eqref{model1} as an equivalent low-rank Tucker decomposition,
\begin{align}
\mathcal{X} &= \cF \times_1 \bU_1 \times_2 \cdots \times_d \bU_d + \mathcal{E}, \quad
\bY = \bU_1 \bF_Y + \bfeta, 
\label{modelpcsc}
\end{align}
where $\bU_i\in\RR^{p_i\times m_i}$ are orthogonal matrices with $m_i\le r_i$. Building on the High Order Orthogonal Iteration (HOOI) framework \citep{zhang2018tensor}, we introduce the Panel Coupled High Order Orthogonal Iteration (PCHOOI) algorithm, which jointly captures the low rank structures present in both $\cX$ and $\bY$. For the coupled first mode, PCHOOI employs a distinctive approach. During initialization, we estimate the leading $m_1$ left singular matrix as $\hbU_1^{(0)}={\rm LSVD}_{m_1} ([\mat_1(\cX),\bY])$. In subsequent iterations, we refine this estimate using the coupled matrix,
$$\hbU_1^{(k)} = {\rm LSVD}_{m_1}([\mat_1(\mathcal{X}\times_2 \hbU_2^{(k-1)\top} \times_3 \cdots \times_d \hbU_d^{(k-1)\top} ), \bY]).$$
For the remaining uncoupled modes, the procedure follows the standard HOOI approach. The complete PCHOOI algorithm is detailed in  Algorithm \ref{PCHOOIalg}.

Following the paradigm of classical spectral clustering, which performs low-rank projection before applying $k$-means \citep{zhang2024leave}, our initialization procedure combines PCHOOI with a $k$-means clustering step. This integrated approach, termed Panel Coupled Matrix-Tensor Spectral Clustering (PMTSC), operates in two stages:

\noindent {\bf Stage 1}: We apply PCHOOI (Algorithm \ref{PCHOOIalg}) to estimate the orthogonal matrices $\bU_i$, obtaining estimates $\hbU_i$ that span the principal subspaces of $\cX$ and $\bY$.

\noindent {\bf Stage 2}: We employ a modified high-order spectral clustering algorithm (Algorithm \ref{PCSCalg}) using these $\hbU_i$ estimates to recover the latent group structures $\hbM_i^{(0)}$. Analogous to PCHOOI, for the coupled first mode, we construct the augmented matrix,
$$\hbZ_1 = \hbU_1\hbU_1^{\top}[ \mat_1(\mathcal{X} \times_2 \hbU_2^{\top} \times_3 \cdots \times_d \hbU_d^{\top}), \bY]. $$
Given the computational complexity of exact $k$-means, we implement a relaxed version that can be efficiently solved using approximation algorithms such as $k$-means++ with relaxation factor $\kappa=O(\log \bar r)$. The pseudo-code is presented in Algorithm \ref{PCSCalg}.

Our algorithm requires the ranks $r_k$ as inputs, which are permitted to scale with tensor dimensions in our theoretical framework. While our simulations assume known true ranks for simplicity, practical applications should employ data-driven selection criteria such as BIC or leverage domain-specific prior knowledge.

\begin{algorithm}[ht]
\caption{Panel Coupled High Order Orthogonal Iteration (PCHOOI)}\label{PCHOOIalg}
\KwIn{ Tensor $\mathcal{X} \in \mathbb{R}^{p_1 \times \cdots \times p_d \times T} $, matrix $\bY \in \mathbb{R}^{p_1 \times T} $, Tucker rank $(m_1, ...,m_d)$, maximum iteration number $K$, tolerance parameter $\epsilon>0$.}
\KwOut{Orthogonal matrices $ \hbU_i = \hbU_i^{(k)}, i = 1,...,d$, tensor $\widehat \cX=\cX\times_{i=1}^d \widehat \bU_i \widehat \bU_i^\top $, and matrix $ \widehat \bY= \widehat \bU_1 \widehat \bU_1^\top \bY$. }
Compute $\hbU_1^{(0)} = {\rm LSVD}_{m_1}([\mat_1(\cX),\bY])$ and $\hbU_i^{(0)} = {\rm LSVD}_{m_i}(\mat_i(\cX))$, $i=2,...,d$.

\For{iterations $k = 1$ \KwTo $K $}{
Compute $\hbU_1^{(k)} = {\rm LSVD}_{m_1}([\mat_1(\mathcal{X}\times_{j=2}^d \hbU_j^{(k-1)\top} ), \bY]). $

\For{$ i= 2$ \KwTo $d$}{
Compute $\hbU_i^{(k)} = {\rm LSVD}_{m_i}(\mat_i( \mathcal{X} \times_{j=1}^{i-1} \hbU_j^{(k)\top} \times_{j=i+1}^d \hbU_{j}^{(k-1)\top}) ) $.
}
\textbf{break if} $\max_{1\le i\le d}\Vert  \hbU_i^{(k)} \hbU_i^{(k)\top} - \hbU_i^{(k-1)} \hbU_i^{(k-1)\top}  \Vert_2^2 \leq  \epsilon $.

}

\end{algorithm}

\begin{algorithm}[htb!]
\caption{Panel Coupled Matrix-Tensor Spectral Clustering (PMTSC)}\label{PCSCalg}
\KwIn{Tensor $\mathcal{X} \in \mathbb{R}^{p_1 \times \cdots \times p_d \times T} $, matrix $\bY \in \mathbb{R}^{p_1 \times T} $, groups number $r_1,..., r_d$, orthogonal matrices $ \hbU_1, ..., \hbU_d$, relaxation factor $\kappa>1$.}
\KwOut{Membership matrices $\hbM_1^{(0)}, ..., \hbM_d^{(0)}$.
}
Calculate $$\hbZ_1 = \hbU_1\hbU_1^{\top}[ \mat_1(\mathcal{X} \times_{j=2}^d \hbU_j^{\top}), \bY] \in \R^{p_1\times (p_{-1}+1)T}, \quad
\hbZ_i = \hbU_i\hbU_i^{\top} \mat_i(\mathcal{X} \times_{j\neq i}^{d} \hbU_j^{\top} ). $$ 
\For{$i = 1$ \KwTo $d$}{
Find $ g_i^{(0)} \in [r_i]^{p_i}$ and centroids $ \hat{s}_1, \ldots, \hat{s}_{r_i} $ such that
$$
\sum_{j=1}^{p_i} \| ( \hbZ_i )_{j:}^\top - \hat{s}_{(g_i^{(0)})_j} \|_2^2 \leq \kappa \min_{\substack{s_1, \ldots, s_{r_i} \\ g_i \in [r_i]^{p_i}}} \sum_{j=1}^{p_i} \| ( \hbZ_i )_{j:}^\top - s_{( g_i)_j} \|_2^2.
$$
}
Construct membership matrix $\hbM_i^{(0)}$ according to $g_i^{ (0)} \in [r_i]^{p_i}$ for all $i=1,...,d$.
\end{algorithm}

\begin{algorithm}[htb!]
\caption{Nearest Neighbor Searching}
\label{NNS}
\KwIn{ Matrix $\hbZ_i^{(k)} $, centroid matrix $\hbC_i^{(k)}  $. }
\KwOut{Membership matrix $\hbM_i^{ (k)}$. 
}
\BlankLine
\For{ j = 1 \KwTo $p_i$ }{Calculate
$
(g_i^{(k)})_j = \arg\min_{a\in [r_i]} \| (  \hbZ_i^{(k)} )_{j:} - ( \hbC_i^{(k)} )_{a:} \|_2^2.
$
}
Construct $\hbM^{(k)}_i$ based on $g_i^{(k)} $ by setting $(\hbM_i^{(k)})_{j,a} = 1$ if and only if $(g_i^{(k)})_j = a$.
\end{algorithm}

\subsubsection{Panel coupled matrix-tensor Lloyd algorithm}

\begin{algorithm}[htb!]
\caption{Panel Coupled Matrix-Tensor Lloyd algorithm (PMTLloyd)}\label{PCLloydalg}
\KwIn{ Tensor $\mathcal{X} \in \mathbb{R}^{p_1 \times \cdots\times p_d \times T} $, matrix $\bY \in \mathbb{R}^{p_1 \times T} $, initial estimate of the membership matrix $\hbM_1^{(0)},..., \hbM_d^{(0)}$,  maximum iteration number $K$.}
\KwOut{Membership matrices $\hbM_i= \hbM_i^{(k)}$, $i=1,...,d$. 
}
\For{  $k = 1 $ \KwTo K}{Compute $\widehat\bP_i^{(k-1)}= \hbM_i^{(k-1)} (\hbM_i^{(k-1)\top}\hbM_i^{(k-1)} )^{-1}$, and let $\widehat\bW_i^{(k-1)}$ be the left singular matrix containing the normalized columns of $\hbM_i^{(k-1)}$, $i=1,...,d$.

Compute $\widehat\cS_i^{(k)}=\cX\times_i \widehat\bP_i^{(k-1)\top} \times_{j\neq i}^d \widehat\bW_j^{(k-1)\top}, \ \bS_Y=\widehat\bP_1^{(k-1)\top}\bY$, $i=1,...,d$.

Calculate 
$\hbC_1^{(k)} = [\mat_1(\widehat\cS_1^{(k)}), \hbS_Y^{(k)}] ,\
\hbZ_1^{(k)} = [\mat_1(\mathcal{X} \times_{j=2}^d \hbW_j^{(k-1)\top}) , \bY ] .$

Apply Algorithm \ref{NNS} $(\hbZ_1^{(k)},\hbC_1^{(k)})$ to get $\hbM_1^{(k)}$.

\For{$i=2$ \KwTo $d$}{
Calculate 
$\hbC_i^{(k)} = \mat_i(\widehat\cS_i^{(k)}), \
\hbZ_i^{(k)} = \mat_i(\mathcal{X} \times_{j=1}^{i-1} \hbW_j^{(k-1)\top} \times_{j=i+1}^d \hbW_j^{(k-1)\top}).$

Apply Algorithm \ref{NNS} $(\hbZ_i^{(k)}, \hbC_i^{(k)})$ to get $\hbM_i^{(k)}$.
}

}

\end{algorithm}

Following warm initialization via PMTSC (Algorithm \ref{PCSCalg}), we employ the Panel Coupled Matrix-Tensor Lloyd algorithm (PMTLloyd, Algorithm \ref{PCLloydalg}) to refine the membership matrices $\bM_i$ and estimate the core tensor $\cS$ and centroid matrix $\bS_Y$. PMTLloyd extends the iterative projection framework developed for Tucker and CP factor models \citep{han2024tensor, han2024cp} to our PMTC model \eqref{model1}. The key insight motivating our approach involves orthogonal projections. Define $\bP_i = \bM_i(\bM_i^{\top} \bM_i)^{-1}$ and consider
\begin{align}
\widetilde\cX_{i}=&\; \cX \times_{j\neq i}^d \bP_j^\top , \quad
\widetilde\cE_{i}=\; \cE \times_{j\neq i}^d \bP_j^\top . 
\end{align}
Since $\bM_i^\top\bP_i=I_{r_i}$, model \eqref{model1} yields
\begin{align} \label{eq:ptmc-ideal0}
\widetilde\cX_{i}=\cS\times_i \bM_i + \widetilde\cE_{i},
\end{align}
where $\widetilde\cX_{i}$ has dimensions $r_1 \times \cdots \times r_{i-1} \times p_i \times r_{i+1} \times \cdots \times r_d$. While this dimension reduction from $p_j$ to $r_j$ (where $r_j\ll p_j$) improves efficiency, the non-orthogonality of $\bP_j,j\neq i$ creates heterogeneous noise in $\widetilde\cE_{i}$. To address this issue, we introduce an orthogonal projection approach. Let $\bW_i$ contain the normalized columns of $\bM_i$, and $\bLambda_i^2=\bM_i^\top \bM_i$. Define
\begin{align}
\cX_{i}=&\; \cX \times_{j\neq i}^d \bW_j^\top , \quad
\cE_{i}^*=\; \cE \times_{j\neq i}^d \bW_j^\top . \label{eq:z}
\end{align}
Then model \eqref{model1} implies that
\begin{align} \label{eq:ptmc-ideal}
\cX_{i}=\cS\times_i \bM_i \times_{j\neq i}^d \bLambda_j + \cE_{i}^*,
\end{align}
Unlike \eqref{eq:ptmc-ideal0}, this formulation preserves signal strength while applying homogeneous noise reduction. Given the true core tensor $\cS\times_{\ell\neq i}^d \bLambda_\ell$ and centroids matrix $\bS_Y$, we estimate $\bM_i$ via nearest neighbor assignment:
\begin{align*}
(g_i)_j &= \arg\min_{a\in [r_1]} \left[\| (\mat_i(\cX_i))_{j:} - (\mat_i(\cS\times_{\ell\neq i}^d \bLambda_\ell))_{a:} \|_2^2 + \| \bY_{j:} - (\bS_Y)_{a:} \|_2^2 \mathbf{1}\{i=1\}  \right].
\end{align*}
The membership matrix is then reconstructed as $(\bM_i)_{j,a} = 1$ if and only if $(g_i)_j = a$, for all $i=1,...,d$.
The operation in \eqref{eq:z} achieves two critical objectives: it dramatically reduces the dimensionality by projecting onto all modes except the $i$-th, and it effectively averages out noise. Under proper conditions on the combined noise tensor $\cE_{i}^*$, estimation of the membership matrix $\bM_i$ based on $\cX_i,\bY$ can be made significantly more accurate, as the statistical error rate now depends on $p_i\prod_{j\neq i}^d r_j$ rather than $p_1p_2\ldots p_d$.

In practice, we do not know $\cS$, $\bS_Y$ and $\bW_{i}$, $1\le i\le d$. Similar to back-fitting algorithms, we iteratively estimate the membership matrix $\bM_{i}$ at iteration $k$ based on
\begin{align} \label{eq:zk}
&\cX_{i}^{(k)}=\; \cX \times_1 \hbW_1^{(k-1)\top} \times_2 \cdots \times_{i-1} \hbW_{i-1}^{(k-1)\top} \times_{i+1} \hbW_{i+1}^{(k-1)\top} \times_{i+2}\cdots\times_d \hbW_{d}^{(k-1)\top},
\end{align}
using estimates $\hbW_{j}^{(k-1)},~ j\neq i$, from the previous iteration. And the centers $\widehat\cS_i^{(k)},\widehat \bS_Y^{(k)}$ are estimated through block-wise averaging and projections $\widehat\cS_i^{(k)}=\cX \times_{i} \widehat\bP_i^{(k-1)\top}\times_{j\neq i}^d \hbW_j^{(k-1)\top}$ and $\widehat\bS_Y^{(k)}=\widehat\bP_1^{(k-1)\top}\bY$.
As we shall show in the next section, such an iterative procedure leads to a much improved statistical rate in the high dimensional panel coupled matrix-tensor clustering scenarios, as if all $\bW_{i}$, $1\le i\le d$, and $\cS,\bS_Y$ are known and we indeed observe $\cX_{i}$ following model \eqref{eq:ptmc-ideal}.

\begin{remark}\label{rmk:Lloyd}
Our orthogonal projection approach shares similarities with Orthogonalized Alternating Least Squares \citep[OALS,][]{sharan2017orthogonalized} for tensor CP decomposition. However, the clustering setting differs fundamentally from CP decomposition. While \cite{tang2025revisit} shows that ALS outperforms OALS in CP decomposition, our framework, based on \eqref{eq:ptmc-ideal}, substantially improves upon approaches using \eqref{eq:ptmc-ideal0}.
The key distinction lies in the singular value structure. Under Assumption \ref{asmp:clusters}, $\lambda_1(\bP_i)\asymp\lambda_{r_i}(\bP_i)\asymp \sqrt{r_i/p_i}$, resulting in greater variability than typical CP base matrices. This heterogeneity degrades the performance of \eqref{eq:ptmc-ideal0}, which underlies the High-order Lloyd algorithm in \cite{han2022exact}. Our simulations in Appendix \ref{section:appendix_coclustering} confirm that our approach consistently outperforms existing methods, even when restricted to clustering solely on $\cX$.

\end{remark}

\subsubsection{Estimation of factor loading matrix}

Let $\widehat \bM_i$ denote the final membership matrix estimates from the PMTLloyd algorithm, and define $\hbW_i=\hbM_i(\hbM_i^\top \hbM_i)^{-1}$. For latent factors, we estimate the factor loading matrix and latent factors as
\begin{align} \label{factor:unobs}
\widehat \bU_B={\rm LSVD}_{m_1}( \hbW_1^\top \bY \bY^\top \hbW_1/T), \quad \widehat \bF=\widehat \bU_B^\top \hbW_1^\top \bY,    
\end{align}
where $\widehat \bU_B$ represents an estimate of the left singular matrix of the loading matrix $\bB$.
As in conventional factor models, this estimate is subject to rotational ambiguity. When factors are observable, we directly estimate the factor loading matrix via least squares method:
\begin{align}\label{factor:obs}
\widehat \bB  = \hbW_1^\top \bY \bF^\top (\bF \bF^\top)^{-1}  .
\end{align}

\section{Theoretical Analysis}
\label{sec:theoretical}

We now establish the statistical consistency and error rates for the proposed estimators and factor loading matrix $\bB$ under proper regularity conditions.
The membership matrix $\bM_i$ encodes cluster assignments through the relationship $(\bM_i)_{j,a} = 1$ if and only if the cluster label $(g_i)_j = a$, for $i=1,...,d$. Define the SVD $\bM_i=\bW_i\bLambda_i\bQ_i^\top$, where $\bW_i$ contains the normalized columns of $\bM_i$, and define the rescaled core tensor as
\begin{align} \label{def:s}
\bS_i=\mat_i(\mathcal{S}\times_{j\neq i}^d \bLambda_j) .     
\end{align}
Because rearranging the cluster labels leaves the clustering outcome unchanged, the cluster label vector $g_i\in\mathbb{R}^{p_i}$ for mode-$i$ can only be determined up to a label permutation. Starting with an initial labeling $g_i^{(0)}\in\mathbb{R}^{p_i}$, we denote by $\pi_i^{(0)} : [r_i] \to [r_i]$ the best permutation that minimizes the discrepancies between $g_i^{(0)}$ and $g_i$, specifically:
\begin{equation}
\pi_i^{(0)} := \arg\min_{\pi \in \Pi_{r_i}} \frac{1}{p_i} \sum_{j=1}^{p_i} \mathbb{I}\left\{(g_i^{(0)})_j \neq (\pi \circ g_i)_j\right\}, 
\end{equation}
where $(\pi \circ g_i)_j := \pi((g_i)_j)$ and $\Pi_{r_i}$ is the collection of all permutations on $[r_i]$.
Let $k$ denote the iteration step in the PMTLloyd algorithm. We define $h_i^{(k)}$ as the mode-$i$ \textit{misclustering error rate}  (CER) at iteration $k$,
\begin{equation} \label{erroreq:h}
h_i^{(k)} := \frac{1}{p_i} \sum_{j=1}^{p_i} \mathbb{I}\left\{(g_i^{(k)})_j \neq (\pi_i^{(0)} \circ g_i)_j\right\}.
\end{equation}
To complement the clustering error rate, we introduce the following misclustering loss,
\begin{align} \label{erroreq:ell}
\ell_i^{(k)} :=& \frac{1}{p_i} \sum_{j=1}^{p_i} \big( \|(\bS_i )_{(g_i^{(k)})_j:} - (\bS_i )_{(\pi_i^{(0)} \circ g_i)_j:} \|_2^2 +  \|(\bS_Y)_{(g_i^{(k)})_j:} - (\bS_Y)_{(\pi_i^{(0)} \circ g_i)_j:} \|_2^2 \mathbf{1}\{i=1\} \big).
\end{align}
We also impose a non-degeneracy condition on the distance between block centers (defined by the core tensor and center matrix) to ensure the identifiability of clustering.
\begin{align}
\Delta_i^2 &:= \min_{j_1 \neq j_2} \left( \|(\bS_i)_{j_1:} - (\bS_i)_{j_2:}\|_2^2 + \|(\bS_Y)_{j_1:} - (\bS_Y)_{j_2:}\|_2^2 \mathbf{1}\{i=1\} \right) > 0, \label{def:delta}\\
\Delta_{i,x}^2 &:= \min_{j_1 \neq j_2} \|(\bS_i)_{j_1:} - (\bS_i)_{j_2:}\|_2^2,\quad  \Delta_{y}^2 := \min_{j_1 \neq j_2} \|(\bS_Y)_{j_1:} - (\bS_Y)_{j_2:}\|_2^2 ,  \label{def:deltax}
\end{align}
for $i=1,...,d$. Then $\Delta_{1}^2 \ge \Delta_{1,x}^2+\Delta_y^2$. Specifically, we define $\Delta_i^2 = \infty$ when $r_i = 1$. Define $\Delta_{\min}=\min_{1\le j\le d} \Delta_j$. Let $p_* = \prod_{i=1}^{d} p_i$, $\bar{p} = \max\{p_1, ...p_d\}$, $\underline{p} = \min\{p_1, ...p_d\} $ and $ p_{-i} = p_*/p_i$. Analogous notation applies for ranks, i.e., $r_*$, $\bar{r}$, $r_{-i}$, and $m_*$, $\bar{m}$, $m_{-i}$.

To present theoretical properties of the proposed procedures, we impose the following assumptions.

\begin{assumption}[Sub-Gaussian noise]
\label{asmp:subgaussian}
Assume $\cE$ is independent of $\cS$ and $\bfeta$ is independent of $\bS_Y$.
Suppose each entry of $\cE$ follows an independent zero-mean sub-Gaussian distribution with sub-Gaussian norm bounded by $\sigma_x$:
\begin{equation}
\E \exp\left(u \cE_{j_1,\ldots,j_d,j_{d+1}}\right) \leq e^{u^2 \sigma_x^2/2}, \quad \forall u \in \mathbb{R}.
\end{equation}
Similarly, suppose each entry of $\bfeta$ follows an independent zero-mean sub-Gaussian distribution with sub-Gaussian norm bounded by $\sigma_y$:
\begin{equation}
\E \exp\left(u \bfeta_{j_1,j_2}\right) \leq e^{u^2 \sigma_y^2/2}, \quad \forall u \in \mathbb{R}.
\end{equation}
\end{assumption}

\begin{assumption}\label{asmp:clusters}
There exists universal positive constants $0 < \alpha < 1 < \beta$ such that
\begin{equation}
\alpha p_i / r_i \leq |\{j \in [p_i] : (g_i)_j = a\}| \leq \beta p_i / r_i, \quad \forall a \in [r_i], \, i \in [d],
\end{equation}
where $|\cdot|$ denotes the cardinality of a given set.
\end{assumption}

\begin{assumption}\label{asmp:mixing}
Let $\bF=(f_{1},...,f_{T})$ with $f_t \in \R^{m_1}$. Assume the factor process $f_{t}$ is stationary and strong $\alpha$-mixing in $t$, with $\E f_{t}=0$. For any $v\in\R^{m_1}$ with $\|v\|_2=1$, 
\begin{align}
& \P\left( \left| v^\top f_{t} \right| \ge x \right) \le c_1 \exp\left( -c_2x^{\gamma_2} \right), \qquad
 c_3\le \E (v^\top f_{t})^2 \le c_4,  \label{cond2}
\end{align}
where $c_1,c_2,c_3,c_4>0$ are constants and $0<\gamma_2\le 2$. In addition, the mixing coefficient satisfies
\begin{align}
\varpi(n) \le \exp\left( - c_0 n^{\gamma_1} \right)  \label{cond1}
\end{align}
for some constant $c_0>0$ and $\gamma_1>0$, where
\begin{align*}
\varpi(n) = \sup_t\Big\{\Big|\P(A\cap B) - \P(A)\P(B)\Big|:
A\in \sigma(f_{s}, s\le t), B\in \sigma(f_{s}, s\ge t+n)\Big\}.
\end{align*}
\end{assumption}

Assumption \ref{asmp:subgaussian} parallels noise conditions commonly imposed in the clustering literature \citep{gao2022iterative, loffler2021optimality, han2022exact}. While we assume independent sub-Gaussian entries for technical convenience and to ensure fast statistical error rates, this condition could theoretically be relaxed to accommodate sub-Gaussian noise with mode-wise additive covariance or Gaussian noise with general cross-sectional dependence. However, such generalizations would substantially complicate the mathematical formulations, statistical results, and technical requirements of our framework. Our choice thus strikes a balance between analytical tractability and the preservation of the core insights of our study.

For analytical tractability, 
Assumption \ref{asmp:clusters} imposes a ``balanced cluster" condition, which ensures that no single group becomes too sparse to allow for consistent recovery of its centroid \citep{gao2022iterative, loffler2021optimality, han2022exact}.

Assumption \ref{asmp:mixing} imposes a standard mixing condition that accommodates a broad class of time series models, including causal ARMA processes with continuously distributed innovations \citep{fan2008nonlinear, tsay2018nonlinear}. This assumption requires the tail probabilities of $f_{t}$ to decay exponentially; notably, when $\gamma_2=2$, the process $f_{t}$ becomes sub-Gaussian.

We begin by analyzing the PCHOOI estimators in Algorithm \ref{PCHOOIalg}, which form the foundation of the PMTSC algorithm (Algorithm \ref{PCSCalg}). Theorem \ref{thm:PCHOOI} establishes performance bounds for the orthogonal loading matrices. This result extends the HOOI framework of \cite{zhang2018tensor} to coupled matrix-tensor analysis and is of independent interest.

\begin{theorem}\label{thm:PCHOOI}
Suppose Assumption \ref{asmp:subgaussian} holds. Define $\lambda_{1,m_1}=\lambda_{m_1}([\mat_1(\cX),\bY])$ and $\lambda_{i,m_i}=\lambda_{m_i}(\mat_i(\cX))$ for $i= 2,...,d$.
Suppose there exists a constant $C_{gap}>0$ that does not depend on $p_i, m_i, \lambda_{i,m_i}$, such that
\begin{align}
\lambda_{1,m_1}& \ge C_{gap}(\sigma_x \sqrt{p_1 + p_{-1}T} + \sigma_y \sqrt{p_1 + T}) , \quad
\lambda_{i,m_i} \ge C_{gap}\sigma_x \sqrt{p_i + p_{-i}T}, \ i= 2,...,d . \label{lambda gap cond}
\end{align}
Then, with probability at least $1- \exp(-c\underline{p} )$, the estimates from Algorithm \ref{PCHOOIalg} satisfy
\begin{align}
&\| \widehat\bU_1 \widehat\bU_1^{\top} - \bU_1 \bU_1^{\top} \|_2 \leq C \frac{\sigma_x \sqrt{p_1 + m_{-1}T} + \sigma_y \sqrt{p_1 + T}}{\lambda_{1,m_1}} , \\
&\| \widehat\bU_i \widehat\bU_i^{\top} - \bU_i \bU_i^{\top} \|_2 \leq C \frac{\sigma_x \sqrt{p_i + m_{-i}T}}{\lambda_{i,m_i}}, \\
&\| \widehat{\cX} - \cX^* \|_F \leq C\sigma_x \sqrt{\bar{p}\bar{m} + m_* T} , \quad 
\| \widehat\bY - \bY^* \|_F \leq C\sigma_y \sqrt{ p_1 m_1 + m_1 T} ,
\end{align}
for $i=2,...,d$.
\end{theorem}

Notably, identifiable cores $\cS,\bS_Y$ in $\cX,\bY$ may exhibit degenerate ranks with $m_i<r_i$. Theorem \ref{thm:PCHOOI} accommodates such degeneracy. While the Frobenius norm error bounds for $\widehat\cX$ and $\widehat\bY$ show no improvement over \cite{zhang2018tensor} through coupled matrix-tensor analysis, the bounds for orthogonal loading matrices $\hbU_i$ reveal important distinctions. For the shared mode 1, the error depends on a weighted average of the spectral norms of both the error tensor and error matrix, scaled by the minimum singular value of the coupled signal matrix. In contrast, for non-shared modes ($i\ge 2$), the error depends solely on the error tensor and the minimum singular values of the signal tensor, consistent with \cite{zhang2018tensor}.

This coupled matrix-tensor analysis provides substantial benefits. In the extreme case where $\bY$ is noiseless ($\sigma_y=0$), the shared component estimation error for mode 1 becomes significantly smaller than in uncoupled analysis for $\cX$, since $\lambda_{m_1}([\mat_1(\cX),\bY])> \lambda_{m_1}(\mat_1(\cX))$. Even with noisy $\bY$, improvement persists: since $\lambda_{m_1}([\mat_1(\cX),\bY])> \lambda_{m_1}(\mat_1(\cX)) + \lambda_{m_1}(\bY)$ typically holds, the statistical error for mode 1 improves compared to uncoupled analysis, provided $\sigma_y\asymp \sigma_x$.

We now establish theoretical guarantees for our clustering algorithms, PMTSC and PMTLloyd. We first present convergence rates for the PMTSC algorithm (Algorithm \ref{PCSCalg}), which serves as initialization for the subsequent PMTLloyd algorithm.

\begin{theorem}[Upper bounds on misclustering rate of PMTSC]
\label{thm:PMTSC}
Suppose Assumptions \ref{asmp:subgaussian} and \ref{asmp:clusters} hold. Let $\kappa>1$. If the SNR satisfies
\begin{align*}
\Delta_1^2 &\ge (C'\kappa r_1/p_1) ( \sigma_x^2(\bar p \bar m + m_* T  )+  \sigma_y^2(p_1 m_1+ m_1 T) ), \quad
\Delta_{i}^2  \ge (C'\kappa r_i/p_i)  \sigma_x^2 (\bar p \bar m + m_* T  ),
\end{align*}
then, with probability at least $1 - C\exp(-c\underline{p})$, the estimates from Algorithm \ref{PCSCalg} satisfy
\begin{align}
\ell_1^{(0)} &\le  \frac{C\kappa }{p_1}  ( \sigma_x^2(\bar p \bar m + m_* T  )+  \sigma_y^2(p_1 m_1+ m_1 T) ) , \quad 
\ell_i^{(0)} \le  \frac{C\kappa }{p_i}   \sigma_x^2 (\bar p \bar m + m_* T  ) ,  \label{thmPMTSC:eq1}\\
h_1^{(0)} &\le \frac{C\kappa ( \sigma_x^2 (\bar p \bar m + m_* T  )+  \sigma_y^2 (p_1 m_1+ m_1 T) ) }{ p_1 \Delta_{1}^2  } ,   \quad
h_i^{(0)} \le \frac{C\kappa \sigma_x^2 (\bar p \bar m + m_* T  ) }{ p_i \Delta_{i}^2  } ,   \label{thmPMTSC:eq2}
\end{align}
for some constant $C,C'>0$, where $i=2,...,d$.
\end{theorem}

Theorem \ref{thm:PMTSC} provides a starting point for our further theoretical analysis. 
It establishes rough upper bounds for both the misclustering error $h_i^{(0)}$ and the loss measure $\ell_i^{(0)}$ ($i=1,...,d$), providing the foundation for our subsequent analysis of the PMTLloyd algorithm. 
While these polynomial rates in Theorem \ref{thm:PMTSC} could be sharpened in the context of the spectral clustering literature, we focus instead on the error bounds of the iterative algorithm, where PMTLloyd substantially improves these initial estimates.

\begin{remark}[Enhanced signal strength]\label{rmk:signal}
Our tensor co-clustering achieves enhanced signal strength compared to Gaussian mixture model clustering algorithms \citep{loffler2021optimality,zhang2024leave,gao2022iterative}. For each mode $i$, the effective signal $\Delta_{i,x}^2$ is amplified by a factor of $c(p_{-i}/r_{-i})$ relative to the minimum distance between rows in $\mat_i(\cS)$, as defined in \eqref{def:s} and \eqref{def:deltax}. This amplification arises from the properties $\lambda_1(\bLambda_i)\asymp\lambda_{r_i}(\bLambda_i)\asymp \sqrt{p_i/r_i}$ according to Assumption \ref{asmp:clusters}. Our coupled matrix-tensor framework further enhances mode 1 by aggregating signals from both the tensor component $\Delta_{1,x}^2$ and the panel matrix component $\Delta_{y}^2$.    
\end{remark}

Next, we examine the statistical performance of the iterative PMTLloyd algorithm (Algorithm \ref{PCLloydalg}) following PMTSC initialization. The dimension reduction operation in \eqref{eq:z} projects $\cX$ in other modes of the tensor from $\R^{p_j}$ to $\R^{r_j}$ for all modes $j\neq i$, preserving cluster centers and signal strength while reducing noise. The following theorem establishes the conditions under which ideal rates, based on population projections, are achieved.

\begin{theorem}[Upper bounds on misclustering rate of PMTLloyd] \label{thm:PMTLloyd}
Suppose Assumptions \ref{asmp:subgaussian} and \ref{asmp:clusters} hold. Let $\{g_i^{(0)}\}_{i=1}^d$ be the initialization of PMTLloyd algorithm and $\{g_i^{(k)}\}_{i=1}^d$ be the estimates at iteration $k$. Assume that for some constants $c>0$, the initialization satisfies
\begin{align}
\ell_i^{(0)} \leq c\min_j\Delta_{j}^2/r_i,
\end{align}
with probability $1-\delta$, and the SNR satisfies
\begin{align}
&\Delta_{1}^2 \gg \sigma_x^2\cdot \frac{r_1 r_* T + \bar p \bar r r_1^2 }{p_1} + \sigma_y^2 \cdot \frac{p_1r_1^2+r_1^2 T}{p_1 }, \quad
\min_{2 \le i\le d} \Delta_{i}^2  \gg \sigma_x^2\cdot \frac{\bar r r_*  T+ \bar p \bar r^2 r_1}{p_{1}}.  \label{thmPMTLloyd:snr1}
\end{align}
Then, with probability at least $1 -\delta- \exp(-c_1\underline{p}) -  \sum_{j=1}^d \exp(- \Delta_{j} ) $, for all $k \geq 1$,
\begin{align}
&\ell_1^{(k)} \leq \exp\left(- \frac{(\frac18-o(1))\Delta_1^{4}  }{ \sigma_x^2 \Delta_{1,x}^{2}+ \sigma_y^2 \Delta_{y}^{2}}  \right)    + \frac{c_2 \Delta_{\min}^2}{2^k}  , \quad 
\ell_i^{(k)} \leq \exp\Big(-\frac{(\frac18-o(1))\Delta_{i}^2}{ \sigma_x^2 }  \Big)  + \frac{c_2 \Delta_{\min}^2}{2^k}  ,   \label{thmPMTLloyd:eq1}\\
&h_1^{(k)} \leq \exp\left(- \frac{(\frac18-o(1))\Delta_1^{4}  }{\sigma_x^2 \Delta_{1,x}^{2}+ \sigma_y^2 \Delta_{y}^{2}}  \right)    + \frac{1}{2^k}  ,  \quad
h_i^{(k)} \leq \exp\Big(-\frac{(\frac18-o(1))\Delta_{i}^2}{ \sigma_x^2} \Big)  + \frac{1}{2^k}  ,  \label{thmPMTLloyd:eq2}
\end{align}
where $c_1,c_2>0$ are constants, $i=2,...,d$. 
\end{theorem}

The SNR requirements in \eqref{thmPMTLloyd:snr1} reveal key differences between coupled and uncoupled modes. For the coupled mode 1, the condition incorporates signals from both the characteristics tensor $\cX$ and panel matrix $\bY$. The first term mirrors the requirement for uncoupled modes, while the second resembles conditions from (sub-)Gaussian mixture model clustering \citep{gao2022iterative, zhang2024leave}. In contrast, uncoupled modes ($i\ge 2$) depend solely on $\cX$.
This signal aggregation provides substantial practical advantages. By Remark \ref{rmk:signal}, when $\sigma_x\asymp \sigma_y$ and $p_i\gg r_i^2$ (as typically holds), the coupled approach achieves weaker SNR requirements than clustering on $\bY$ alone. Similarly, when $\Delta_y$ is large, the requirements are weaker than clustering on $\cX$ alone. Thus, coupling never strengthens SNR requirements and often relaxes them considerably.

\begin{remark}[Improved misclustering error bounds]\label{rmk:cerbdd}
For coupled mode 1 with $\sigma_x=\sigma_y$, the misclustering error bound \eqref{thmPMTLloyd:eq2} simplifies to:
\begin{align*}
h_1^{(k)} \leq \exp\left(- \frac{(\frac18-o(1))\Delta_1^{4}  }{\sigma_x^2 \Delta_{1,x}^{2}+ \sigma_y^2 \Delta_{y}^{2}}  \right)     + \frac{1}{2^k}   =\exp\left(- \frac{(\frac18-o(1))\Delta_{1,x}^{2}  }{\sigma_x^2}  \right) \exp\left(- \frac{(\frac18-o(1))\Delta_{y}^{2}  }{\sigma_y^2}  \right)    + \frac{1}{2^k} . 
\end{align*}
This bound equals the product of optimal misclustering rates for $\cX$ and $\bY$ separately, guaranteeing improvement over single-source clustering. Both simulations and empirical analyses confirm that with properly chosen weight $\omega$ in \eqref{loss2}, the coupled framework uniformly outperforms clustering based solely on $\cX$ or $\bY$.

While the error bound \eqref{thmPMTLloyd:eq2} for uncoupled modes ($i\ge2$) appears independent of mode 1 coupling---matching tensor co-clustering bounds---our simulations reveal improvements even in these modes under moderate SNR. This likely stems from improved $\bM_1$ estimation, which enhances projection updates in \eqref{eq:zk} and propagates benefits throughout the algorithm.

For uncoupled modes, \eqref{thmPMTLloyd:eq2} achieves the optimal constant $1/8$ in the exponent, improving upon \cite{han2022exact}. Even when restricted to clustering solely on $\cX$, our PMTLloyd algorithm outperforms existing methods (see Remark \ref{rmk:Lloyd}), with empirical validation provided in Appendix~\ref{section:appendix_coclustering}.

\end{remark}

Combining Theorems \ref{thm:PMTSC} and \ref{thm:PMTLloyd} yields the following exact clustering results.

\begin{corollary}\label{thm:exact_recovery}
Let $\{g_i^{(k)}\}_{i=1}^d$ denote the membership vectors at iteration $k$ of the PMTLloyd algorithm, with $\{g_i^{(0)}\}_{i=1}^d$ being the output of the PMTSC algorithm. Under the conditions of Theorem \ref{thm:PMTLloyd}, for some constant $c_1>0$, with probability at least $1 - \exp(-c_1\underline{p}) -  \sum_{j=1}^d \exp(- \Delta_{j} ) $, we achieve exact clustering of $\{g_i\}_{i=1}^d$ when $K \geq 2\lceil\log\bar{p}\rceil$. That is, there exist permutations $\{\pi_i\}_{i=1}^d$ such that
\begin{equation}
g_i^{(K)} = \pi_i \circ g_i, \quad \text{for all } i = 1, \ldots, d.
\end{equation}
\end{corollary}

Beyond recovering cluster memberships in the PMTC model \eqref{model1}, another important task is estimating the factor loading matrix $\bB$. We establish guarantees for the estimated loading matrix under both observable and unobservable factor scenarios.

\begin{theorem}[Factor Loading Estimation] \label{thm:factors}
Suppose Assumptions \ref{asmp:subgaussian}, \ref{asmp:clusters}, and \ref{asmp:mixing} hold. Let $\hbM_i$, $1\le i\le d$ denote the membership matrices obtained from the PMTLloyd algorithm under the conditions of Theorem \ref{thm:PMTLloyd}.

(i) Latent factors. When the factor process $\bF$ is latent, let $\bB$ have SVD $\bB=\bU_B \Lambda_B \bV_B^\top$ and $\lambda_B=\lambda_{\min}(\bB) \asymp \lambda_{\max}(\bB)$. Assume $m_1=O(T)$. Then, with probability at least $1 - e^{-c_1(\log (T\wedge r_1)+r_1)} - \exp(-c_1\underline{p}) -   \sum_{j=1}^d \exp(- \Delta_{j} ) $, the estimated factor loading matrix $\widehat\bB$ from \eqref{factor:unobs} satisfies
\begin{align}\label{thmfactors:eq1}
 \| \widehat \bU_B \widehat \bU_B^\top - \bU_B \bU_B^\top\|_2 \le   c_2  \frac{ \sigma_y }{\lambda_B} \sqrt{\frac{r_1(r_1+\log(T\wedge r_1) ) }{Tp_1}}  +  c_2  \frac{ \sigma_y^2}{\lambda_B^2}  \sqrt{\frac{r_1^2(r_1+\log(T\wedge r_1)) }{T p_1^2}},
\end{align}
where $c_1,c_2>0$ are constants. 

(ii) Observed factors. When the factor process $\bF$ is observable, with probability at least $1 - T^{-c_1}-\exp(-c_1\underline{p}) -  \sum_{j=1}^d \exp(- \Delta_{j} ) $, the estimated factor loading matrix $\widehat\bB$ from \eqref{factor:obs} satisfies
\begin{align}
\| \widehat \bB - \bB \|_F \le   c_2 \sigma_y   \sqrt{\frac{m_1 r_1^2 \log(r_1T)}{Tp_1}}  , \quad 
\max_{1\le i\le r_1} \| \widehat \bb_{i} - \bb_i \|_2 \le   c_2 \sigma_y   \sqrt{\frac{m_1 r_1 \log(r_1T)}{Tp_1}}  , \label{thmfactors:eq2}
\end{align}
where $c_1,c_2>0$ are constants, and $\bb_i$ is the $i$-th row of $\bB$.
\end{theorem}

In the latent factors case, the factor loading matrix exhibits rotational ambiguity, requiring that estimation accuracy be measured by the distance between column subspaces, as is standard in the latent factor literature  \citep{bai2003,lam2012,han2024tensor}.
Remarkably, consistency of $\widehat\bU_B$ holds even with finite or slowly growing $T$, provided weak factor strength $\lambda_B\asymp \sigma_y$ and dimensionality requirement $p_1\gg r_1^2$. A stronger factor strength further relaxes the dimensionality requirement for finite $T$.
Similarly, for observed factors, consistency requires only slowly growing $T$ when $p_1\gg m_1r_1$. Intuitively, the grouping structure effectively provides $p_1/r_1$ repeated samples per cluster pattern, reducing noise and weakening requirements on $T$. This is a unique advantage of grouped panel data.

\begin{remark}[Comparison with ungrouped factor analysis]
Our grouped approach yields substantial improvements over standard factor analysis. For latent factors with strong factor strength $\lambda_B \asymp \sqrt{r_1}\sigma_y$, let $\widehat\bU_1$ and $\bU_1$ denote the top $m_1$ left singular matrix of $\widehat\bM_1 \widehat\bU_B$ and $\bM_1 \bU_B$, respectively. Then bound \eqref{thmfactors:eq1} leads to $\| \widehat \bU_1 \widehat \bU_1^\top - \bU_1 \bU_1^\top\|_2 = O(\sqrt{r_1/(Tp_1)})$, compared to $O(\sqrt{1/T})$ without grouping structure \citep{lam2012}. For observed factors, standard ungrouped analysis achieves $\max_i\| \widehat \bb_{i} - \bb_i \|_2 = O(\sqrt{m_1\log(p_1)/T})$ \citep{fan2011high}, substantially slower than our rate in \eqref{thmfactors:eq2}. This also improves upon existing grouped panel regression rates \citep{su2016identifying}.

\end{remark}

\section{Simulation Studies}
\label{sec:simulation}

We evaluate the clustering recovery and loading matrix estimation accuracy of our methods via simulations under two data-generating processes.
By varying parameters such as SNR, we benchmark our coupled approach against methods using only the tensor $\cX$ or the matrix $\bY$. These experiments demonstrate how joint estimation enhances clustering assignments, factor loading estimation, and reconstruction accuracy.

\subsection{Simulation for PCHOOI Algorithm}

In this subsection, we evaluate the proposed PCHOOI algorithm under model \eqref{modelpcsc} with $d=2$. The orthogonal loading matrices $\bU_i$ ($i=1,2$) are obtained as the left singular matrix of $p_i \times m_i$ matrices with i.i.d.\ $\mathcal{N}(0,1)$ entries. The noise tensor $\mathcal{E} \in \mathbb{R}^{p_1 \times p_2 \times T}$ and error matrix $\bfeta \in \mathbb{R}^{p_1 \times T}$ have i.i.d.\ entries from $\mathcal{N}(0, \sigma_x^2)$ and $\mathcal{N}(0, \sigma_y^2)$, respectively.
We set $p_1 = p_2 = 50$, $T = 40$, $\sigma_x = \sigma_y = 1$, $m_1 = m_2 = 5$, $\lambda_{\min}(\mathcal{S}) = c_x \sqrt{p_1 + m_*T}$, and $\lambda_{\min}(\bS_Y) = c_y \sqrt{p_1 + T}$. Two scenarios are considered: (i) $\log c_x = 1$ with $\log c_y$ varying from 1 to 3; and (ii) $\log c_y = 2$ with $\log c_x$ varying from 0 to 2. We compare Algorithm \ref{PCHOOIalg} against SVD on $\bY$ and HOOI on $\cX$, reporting average performance over 100 replications. Performance is measured by the subspace distance $\ell_2(\bU_i, \hbU_i) = \|\hbU_i \hbU_i^\top - \bU_i \bU_i^\top\|_2$, $i=1,2$.

\begin{figure}[htb!]
\caption{\footnotesize \textbf{Estimation errors $\ell_2(\hbU_1,\bU_1)$ and $\ell_2(\hbU_2,\bU_2)$}}
\vspace{-20pt}
 \begin{center}
\adjustbox{center}{\includegraphics[width=1\textwidth]{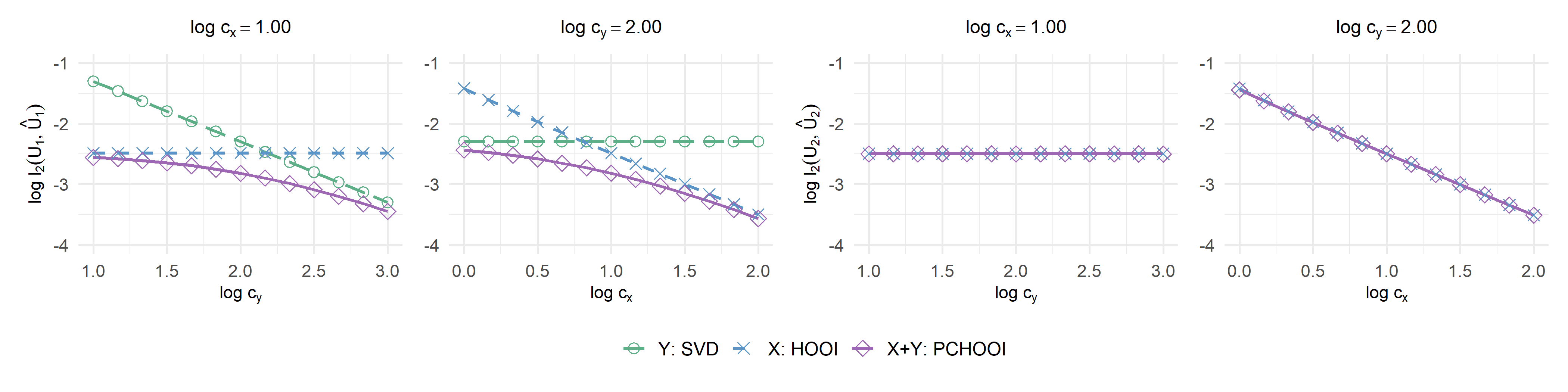}}   
 \end{center}
\vspace{-20pt}
\footnotesize{\underline{Notes:} The left two panels show errors for $\hbU_1$; the right two panels show errors for $\hbU_2$. Results are averaged over 100 repetitions.}
\vspace{-10pt}
\label{fig-PCHOOI}
\end{figure}

By leveraging information from both $\bY$ and $\cX$, PCHOOI delivers more accurate loading space estimates, particularly for the shared first mode. 
For the coupled mode ($\bU_1$), PCHOOI achieves accuracy exceeding that of the better-performing baseline, with the largest gains occurring when SVD and HOOI perform similarly. In contrast, for the uncoupled mode ($\bU_2$), improvements are minimal, with $\ell_2(\bU_2, \hbU_2)$ decreasing by less than 0.1\%. 
This results from the theoretical structure of the PCHOOI estimator; coupling directly increases the singular values for the shared mode but only provides secondary benefits to uncoupled modes.

\subsection{Simulation for Clustering and Factor Loading Estimation}
In this subsection, we evaluate the proposed clustering algorithms under model \eqref{model1} with $d=2$. The noise tensor $\mathcal{E}$ has independent entries from a zero-mean sub-Gaussian distribution with sub-Gaussian norm $\sigma_x$, and the error matrix $\eta$ has independent entries from a zero-mean sub-Gaussian distribution with sub-Gaussian norm $\sigma_y$. Group assignments are balanced across clusters, where each entity has an equal probability of being assigned to any cluster.

\begin{figure}[htb!]
\caption{\footnotesize \textbf{Clustering Error Rate (CER) for different methods} }
\vspace{-20pt}
\begin{center}
\adjustbox{center}
{\includegraphics[width=1\textwidth]{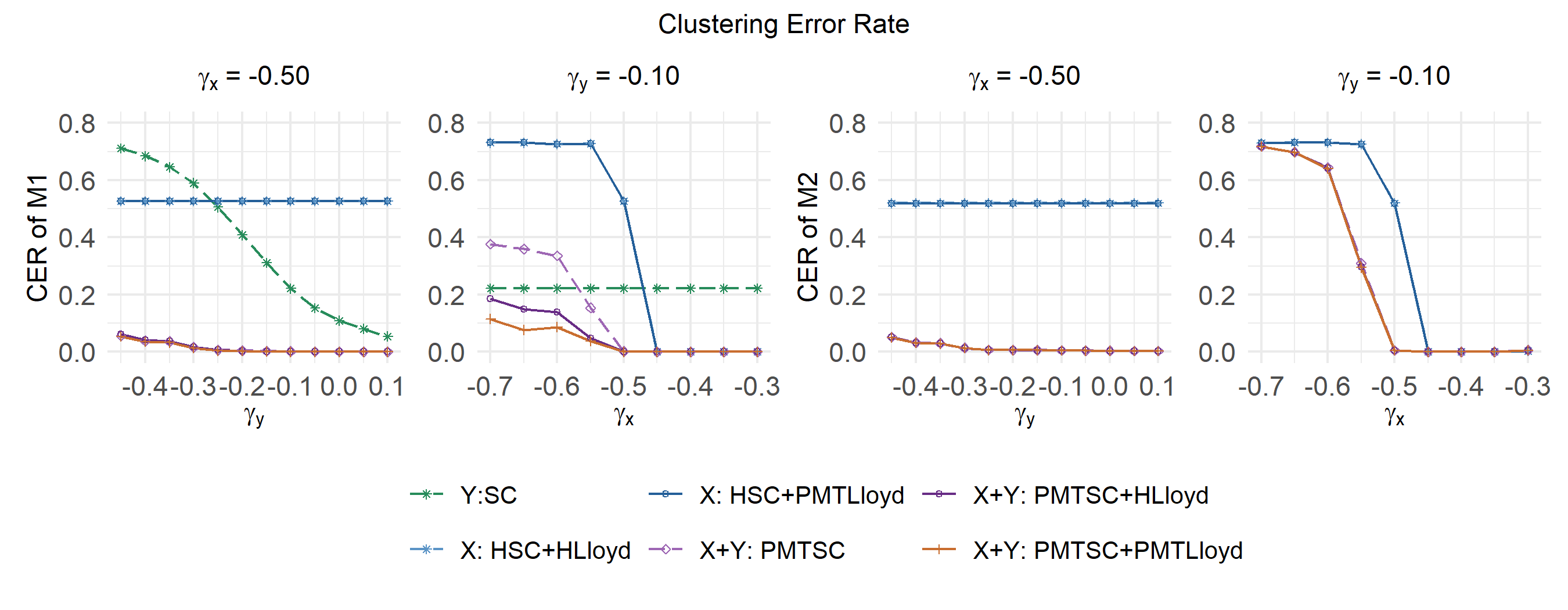}}    
\end{center}
\vspace{-25pt}
\footnotesize{\underline{Note:} The left two panels display the CER of the first mode (coupled); the right two panels show the CER of the second mode (uncoupled). The first and third panels fix $\gamma_x = -0.5$ and vary $\gamma_y$ from $-0.45$ to $0.1$; the second and fourth panels fix $\gamma_y = -0.10$ and vary $\gamma_x$ from $-0.7$ to $-0.3$. 
Results are averaged over 100 repetitions.}
\vspace{-10pt}
\label{fig_CER_PMTC}
\end{figure}

For the core tensor, factor loading matrix, and factors, we set $\mathcal{S}_{j_1j_2t} \sim \mathcal{N}(0, \sigma_s^2)$ with $\mathcal{S} \in \mathbb{R}^{r_1 \times r_2 \times T}$, $\bb_{i } \sim \mathcal{N}(\mu_B, \sigma_B^2 \bI_{m_1})$ for $i=1,...,r_1$ where $\bb_i$ is the $j$-th row of $\bB$, and $f_t \sim \mathcal{N}(\mu_f, \sigma_f^2 \bI_{m_1})$ with $f_t \in \mathbb{R}^{m_1 }$, where $\bI_{m_1}$ denotes the $m_1 \times m_1$ identity matrix. Throughout, we set $r_1 = r_2 = m_1 = 5$, $p_1 = p_2 = 200$, $T = 120$ and $\sigma_x = \sigma_y = \sigma_s = \sigma_B = \sigma_f = 1$. We specify $\mu_B = (1, 1, 1, 0, 0)$ and $\mu_f = 0.03 \cdot \mathbf{1}_5$, where $\mathbf{1}_5$ denotes the 5-dimensional vector of ones. After generating $\mathcal{S}$ and $\bB$, we normalize their magnitudes to satisfy the following SNR constraints motivated by \eqref{thmPMTLloyd:snr1}:
\begin{align}
{\rm SNR}_{x} &= \frac{\Delta_{x } ^2 }{ \sigma_{x } ^2 } 
=C_x (T + \bar{p})p_*^{\gamma_{x} } , \quad 
{\rm SNR}_{y } = \frac{\Delta_{y } ^2 }{ \sigma_{y } ^2 } 
= C_y\frac{(T + \bar{p}) p_*^{\gamma_{y } }}{ r_2} ,
\end{align}
where $\bar{p} = \max\{p_1, p_2\}$, $p_* = p_1 p_2$, and $\Delta_{x}^2 = \min_{i} \Delta_{i,x}^2 $ with $\Delta_{i,x}$ and $\Delta_{y}$ defined in \eqref{def:deltax}. We consider two scenarios: (i) fix $\gamma_x = -0.50$ and vary $\gamma_y$ from $-0.45$ to $0.10$; (ii) fix $\gamma_y = -0.10$ and vary $\gamma_x$ from $-0.7$ to $-0.3$.

We compare our coupled algorithms, PMTSC (``X+Y: PMTSC'') and PMTLloyd refinement with PMTSC initialization (``X+Y: PMTSC+PMTLloyd''), against several benchmarks that utilize $\cX$ and/or $\bY$. For convenience, HLloyd denotes the High-order Lloyd algorithm representing the projection idea in \eqref{eq:ptmc-ideal0}, applicable to coupled $\cX, \bY$, or $\cX$ alone; similarly, PMTLloyd represents the orthogonal projection in \eqref{eq:ptmc-ideal} and can also be applied solely to $\cX$. The benchmarks are: spectral clustering on $\bY$ \citep[``Y: SC'';][]{zhang2024leave}, HLloyd refinement with High-order Spectral Clustering (HSC) initialization on $\cX$ \citep[``X: HSC+HLloyd'';][]{han2022exact}, PMTLloyd refinement with HSC initialization on $\cX$ (``X: HSC+PMTLloyd''), and HLloyd refinement with PMTSC initialization on coupled $\cX, \bY$ (``X+Y: PMTSC+HLloyd''). As expected, increasing $\gamma_x$ or $\gamma_y$, which corresponds to higher ${\rm SNR}_x$ and ${\rm SNR}_y$, reduces the Clustering Error Rate (CER) for all methods.

The left two panels of Figure \ref{fig_CER_PMTC} show clustering accuracy for the first mode (coupled mode). Methods based on coupled $\cX, \bY$ achieve uniformly lower CER than methods using $\cX$ alone, and also outperform methods using only $\bY$, except for PMTSC; PMTSC+PMTLloyd is consistently the best performer. 
In the second panel with varying $\gamma_x$, on coupled $\cX,\bY$, PMTSC+HLloyd improves over PMTSC, while PMTSC+PMTLloyd further improves over PMTSC+HLloyd, demonstrating both the benefit of orthogonal projection and the advantage of iterative refinement. The right two panels show clustering accuracy for the second mode (uncoupled mode). Surprisingly, methods based on coupled $\cX, \bY$ still achieve uniformly lower CER than methods using $\cX$ alone. This likely stems from improved first-mode clustering, which enhances the projection updates in \eqref{eq:zk} and propagates benefits throughout the algorithm. In the first, third, and fourth panels, all coupled methods coincide. 
Although PMTLloyd and HLloyd applied solely to $\cX$ are indistinguishable in these figures, Appendix \ref{section:appendix_coclustering} demonstrates the advantages of PMTLloyd over HLloyd in tensor co-clustering, particularly with imbalanced clusters.
These simulation results demonstrate the consistent superiority of coupled clustering methods and the benefits of PMTLloyd refinement.

We next evaluate the accuracy of estimated factor loadings. For observable factors, we compute $\sqrt{\sum_{i=1}^{p_1}\| \widehat\bb_i - \bb_i \|_{2}^2}$ as the estimation error, applying a time-series demeaning step to $\bY$ before estimation. For latent factors, we measure accuracy using the subspace distance $\| \widehat \bU_B \widehat \bU_B^{\top} - \bU_B \bU_B^{\top} \|_2$, where $\hbU_B = \text{LSVD}_{m_1}(\hbM_1 \hbB)$ and $\bU_B = \text{LSVD}_{m_1}(\bM_1 \bB)$.

\begin{figure}[htb!]
\caption{ \textbf{Factor loading estimation error for different methods} }
\vspace{-20pt}
\begin{center}
\adjustbox{center}{\includegraphics[width=1\textwidth]{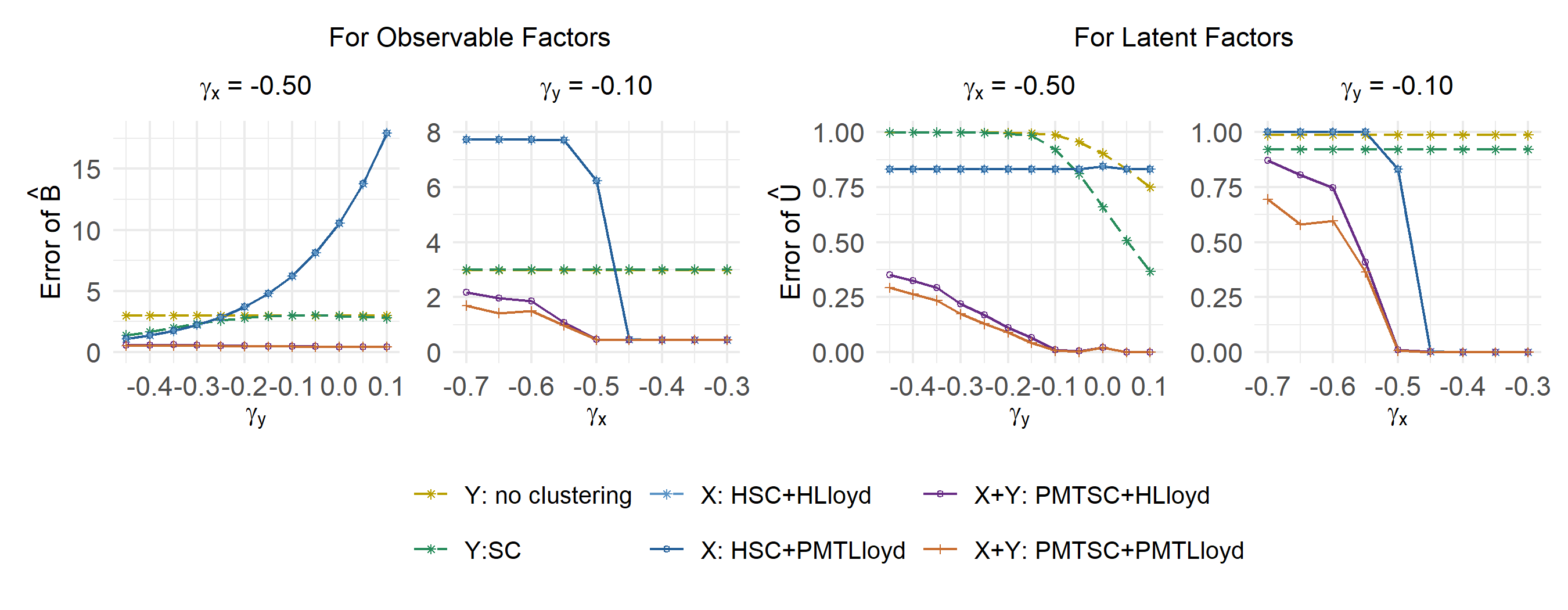}}  
\end{center}

\vspace{-25pt}
\footnotesize{\underline{Note:}
The left two panels display errors for the observable factors case; the right two panels show errors for the latent factors case. The first and third panels fix $\gamma_x = -0.5$ and vary $\gamma_y$ from $-0.45$ to $0.1$; the second and fourth panels fix $\gamma_y = -0.10$ and vary $\gamma_x$ from $-0.7$ to $-0.3$. Results are averaged over 100 repetitions.
}
\vspace{-10pt}
\label{fig-B-err}
\end{figure}

Figure \ref{fig-B-err} displays the factor loading matrix estimation errors. The superiority of coupled methods over those using $\cX$ or $\bY$ alone is reflected in the clustering results, where PMTSC+PMTLloyd consistently outperforms PMTSC+HLloyd, remaining the best performer. For observable factors, when $\gamma_y$ is very small (low SNR in $\bY$), all clustering algorithms outperform no-clustering estimation, even with high CER. However, as $\gamma_y$ increases, $\cX$-based methods deteriorate due to clustering errors, while coupled methods avoid this degradation. Although spectral clustering on $\bY$ eventually improves as its CER approaches zero, our coupled methods converge faster and maintain the best performance throughout.

Appendix \ref{section:appendix_balance} presents additional SNR settings under balanced clusters, Appendix \ref{section:appendix_imbalance} extends to imbalanced clusters, and Appendix \ref{section:appendix_small} addresses smaller dimensions. Across all settings, we observe the same phenomena. Overall, coupling $\mathcal{X}$ with $\bY$ yields significantly lower clustering errors and superior factor loading matrix estimation.

\section{Applications to Empirical Asset Pricing}
\label{sec:application}

\paragraph{Data.}
We apply our methodology to U.S. equity portfolios constructed via the Panel Tree (P-tree) approach \citep*{cong2025growing}.
This flexible method allows for portfolio construction based on multiple asset characteristics, capturing nonlinear and interactive effects.
The dataset comprises monthly returns and 61 characteristics for 400 portfolios spanning from January 1990 to December 2024. From this data, we construct a characteristics tensor $\mathcal{X}$ and a return matrix $\bY$. To ensure comparability across firms, all characteristics within $\mathcal{X}$ are rank-normalized on a cross-sectional basis.

For modeling observable factors, we employ the Fama-French five-factor model, given its established relevance in explaining cross-sectional variations in stock returns. 
The five factors are the excess market return (Mkt-RF), size (SMB), value (HML), profitability (RMW), and investment (CMA).
This allows us to benchmark the performance of our proposed methodology against a widely accepted framework in empirical asset pricing.

\paragraph{Empirical Design.}
We set the number of clusters in the first mode to $r_1 \in \{2, 5, 10, 25\}$, while the second mode is fixed at $r_2 = 6$. This choice aligns with the widely recognized clustering of stock characteristics into six distinct themes: momentum, value, investment, profitability, frictions related to size, and intangibles.

To assess model performance, we use the total $R^2$, a standard metric for evaluating cross-sectional model fit \citep[e.g.,][]{feng2024deep}. This measure captures the proportion of return variation explained by the factor model relative to a market benchmark:
\[
\text{total } R^2 = 1 - \frac{\sum_{i=1}^{p_1} \sum_{t=1}^T \left( Y_{i,t} - \sum_{k=1}^{r_1} \hat{\bb}_{k}^\top f_t \cdot \mathbf{1}\{i \in \mathcal{G}_{1k}\} \right)^2}{\sum_{i=1}^{p_1} \sum_{t=1}^T \left( Y_{i,t} - R_t^{\text{mkt-rf}} \right)^2},
\]
where $Y_{i,t}$ represents observed returns for asset $i$ at time $t$, $\hat{\bb}_{k}$ are estimated factor loadings for group $k$, $f_t$ denotes factor realizations, and $R_t^{\text{mkt-rf}}$ represents excess market returns. 
The denominator benchmarks the variation explained by the market factor. A positive $R^2$ signifies the factor model's capacity to explain additional variation.

We benchmark PMTC model \eqref{model1} against a range of alternatives, including univariate sorts (e.g., book-to-market ratio ``BM" and market equity value ``ME"), $5 \times 5$ bivariate sorts, and economically motivated specifications with pre-defined clusters on the second mode ($\mathcal{G}_2^F$). These comparisons evaluate PMTC relative to return-based baselines, widely used sorting methods, and economically grounded groupings.

For data splitting, we employ two evaluation schemes to assess our methodology. The first approach uses an in-sample (INS) and out-of-sample (OOS) split, where the dataset is divided into a training sample covering the first 35 years (January 1980-December 2014) and an OOS validation period spanning the subsequent 10 years (January 2015-December 2024). 
The training sample is used to estimate the latent clustering structures ($\hbM_1$ and $\hbM_2$) and the factor loading matrix ($\hbB$). 
Predictive performance is then evaluated on the OOS period. Under this design, the algorithm inputs are $\mathcal{X}_{\text{train}} \in \mathbb{R}^{400 \times 61 \times 420}$, $\bY_{\text{train}} \in \mathbb{R}^{400 \times 420}$, and $\bF_{\text{train}} \in \mathbb{R}^{5 \times 420}$. An alternative scheme is provided in Appendix \ref{section:appendix_data}.

\begin{table}[htb!]
\centering

\label{tab2}
\captionsetup[table]{skip=6pt}
\resizebox{\ifdim\width>\linewidth\linewidth\else\width\fi}{!}{
\begin{threeparttable}
\caption{\textbf{Empirical Comparison of Methods}}
\footnotesize
\begin{tabular}{lcccccccccccc}
\toprule
& \multicolumn{6}{c}{In-Sample (Train) $R^2$ (\%)} & \multicolumn{6}{c}{Out-of-Sample (Validation) $R^2$ (\%)}\\
\cmidrule(lr){2-7}\cmidrule(lr){8-13}
& \multicolumn{3}{c}{Benchmarks} & \multicolumn{3}{c}{\textbf{Our methods}} 
& \multicolumn{3}{c}{Benchmarks} & \multicolumn{3}{c}{\textbf{Our methods}}\\
\cmidrule(lr){2-4}\cmidrule(lr){5-7}\cmidrule(lr){8-10}\cmidrule(lr){11-13}
$r_1$  & Only $\bY$ & BM & ME & $\mathcal{G}_2^{F}$ & PMTSC & PMTLloyd
       & Only $\bY$ & BM & ME & $\mathcal{G}_2^{F}$ & PMTSC & PMTLloyd\\
\midrule
& \multicolumn{12}{c}{1980--2014 (INS) / 2015--2024 (OOS) Split}\\
\cmidrule(lr){5-11}
1  & 16.3 &      &      &      &      &      
   & 17.6 &      &      &      &      &      \\
2  & 28.2 &      &      & \textbf{29.8} & 26.2 & 29.3
   & 24.7 &      &      & 26.6 & 25.4 & \textbf{26.9} \\
5  & 31.7 & 18.3 & 18.5 & 34.5 & 30.8 & \textbf{34.7}
   & 27.7 & 21.7 & 21.5 & 30.4 & 28.2 & \textbf{30.6} \\
10 & 31.5 & 18.8 & 27.6 & \textbf{36.6} & 32.1 & 36.3
   & 27.9 & 21.8 & 26.0 & 30.9 & 28.9 & \textbf{31.8} \\
25 & 33.3 & \multicolumn{2}{c}{29.5} & 37.4 & 33.6 & \textbf{37.5}
   & 28.6 & \multicolumn{2}{c}{28.2} & \textbf{32.6} & 29.5 & \textbf{32.6}  \\
\bottomrule
\end{tabular}
\footnotesize{ \underline{Note:} 
This table reports cross-sectional $ R^2 $ (\%) for training (In-Sample, INS) and validation (Out-of-Sample, OOS) data across methods. The first three columns (Benchmarks) show baseline approaches: ``Only $Y$'' clusters on the returns matrix $\bY$, while ``BM'' and ``ME'' use univariate sorts with joint characteristic benchmarks. The last three columns (Ours) present our methods. ``$ \mathcal{G}_2^{F} $'' employs a pre-defined economic grouping for model-2, and ``PMTSC'' and ``PMTLloyd'' fix the second-mode cluster count at $r_2=6$. The table utilizes 35 years of data (1980-2014) for training and evaluates the model on the years 2015-2024. These methods aim to enhance clustering accuracy by incorporating economic structure.}
\end{threeparttable}
}
\vspace{-10pt}
\end{table}

\paragraph{Empirical Performance Comparison.}
PMTLloyd consistently outperforms competing methodologies in both in-sample (INS) and out-of-sample (OOS) evaluations.
At $r_1 = 10$, a standard benchmark in empirical asset pricing \citep[e.g.,][]{Fama1992crosssection}, it enhances the OOS total $R^2$ by 2.9 percentage points compared to the ``Only $\bY$" baseline under a simple in-sample and out-of-sample split. 

Relative to traditional sorting methods, PMTLloyd demonstrates considerable improvements. At $r_1 = 10$, the OOS total $R^2$ is 10 percentage points higher than BM and 5.8 percentage points higher than ME. 
This highlights its capacity to outperform standard approaches across both static and dynamic settings.
When compared to the $\mathcal{G}_2^F$ specification, which relies on predefined characteristic clusters, PMTLloyd generally achieves comparable or superior performance, further validating its adaptability and effectiveness in diverse settings.

\begin{figure}[htb!]
 \caption{\textbf{The coefficient of factors in different groups}}
 \vspace{-20pt}
\begin{center}
\begin{minipage}[htb!]{0.47\textwidth}
\centering
\includegraphics[width=1\textwidth]{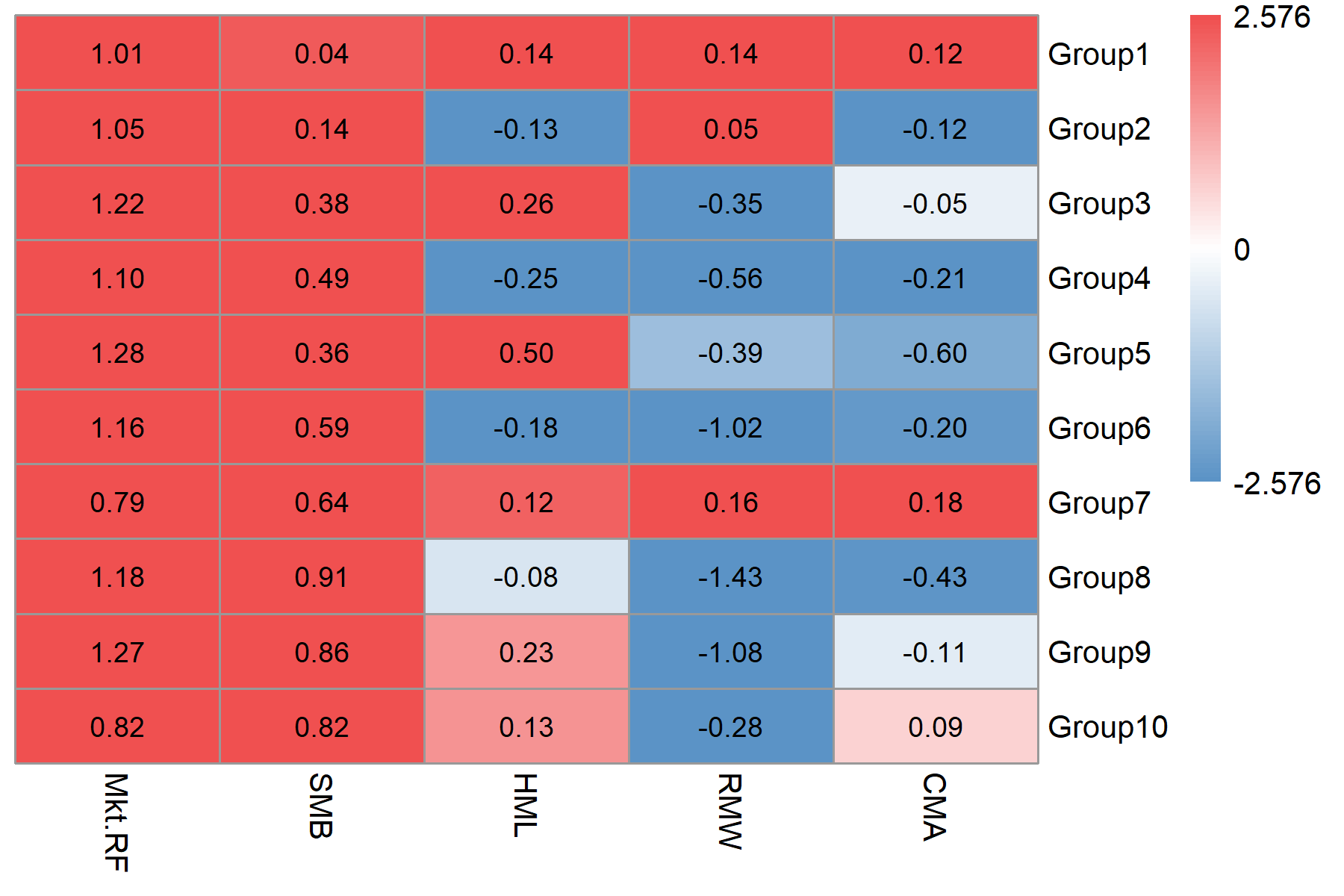}
\end{minipage}
\begin{minipage}[htb!]{0.47\textwidth}
\centering
\includegraphics[width=1\textwidth]{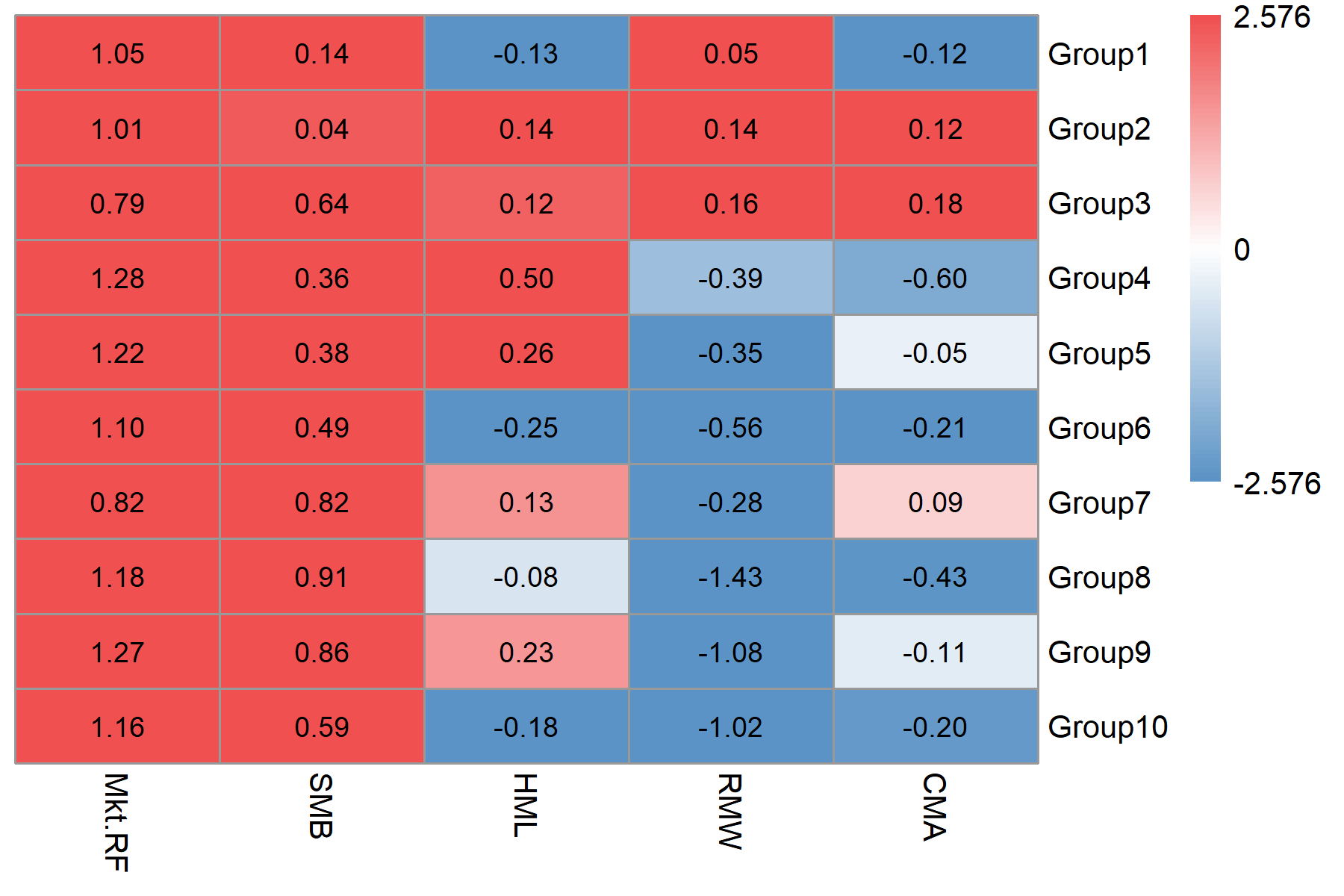}
\end{minipage}   
\end{center}
\footnotesize{\underline{Note:} This figure presents estimated coefficients from the Fama-French five-factor (FF5) model for ten portfolios, sorted by average market equity (ME) in the left panel and profit margin (PM) in the right. Each cell shows factor loadings, indicating exposure to each factor. The color scale reflects Newey-West t-statistics: red for significant positive exposures, blue for significant negative exposures, and white for near-zero coefficients. T-statistics are capped at $[-2.576, 2.576]$, corresponding to a 1\% significance level.

}
\vspace{-10pt}
    \label{fig13}
\end{figure}

Using a standard INS-OOS split, we classify the $r_1 = 10$ clusters by ranking them in descending order based on within-group average market equity (ME) and profit margin (PM). 
This methodology underscores significant cross-group heterogeneity. 
Clusters characterized by high ME and PM exhibit market betas close to 1 and positive RMW loadings, suggesting that larger firms not only tend to co-move with the market but also demonstrate stronger profitability.

The left panel illustrates a monotonic increase in SMB (size) exposure as market equity declines, while Groups 7 and 8 -- characterized by the highest SMB betas -- simultaneously exhibit the most negative RMW (profitability) loadings.
The right panel shows a consistent decline in RMW betas as PM decreases, aligning with the definition of the profitability factor. This trend highlights the systematic reduction in RMW beta with lower PM values.
Moreover, clusters with similar market, SMB, and RMW betas often exhibit substantial differences in their HML and CMA loadings, highlighting variation in value and investment characteristics across groups.
The systematic alignment between latent clusters and fundamental characteristics, such as market equity and profit margins, indicates that PMTC effectively captures the low-rank structure of risk premia, leading to the observed gains in $R^2$.

\section{Final Discussion}
\label{sec:discussion}
While this study focuses on a panel matrix-tensor setting with no time-mode clustering, our framework naturally extends to general cases that relax this restriction.
Coupling a matrix and a tensor, even when they share only a single-mode grouping structure, can still be highly informative: the shared mode stabilizes estimation and, in practice, improves recovery of latent clusters in other modes. This flexibility of our approach suggests that the benefits of joint modeling extend beyond the panel setup considered here. More broadly, settings with multiple modes sharing a common grouping structure could further enhance the precision of structure recovery.

Another natural statistical extension of our framework arises when the group structures in $\mathcal{X}$ and $\bY$ are not perfectly aligned. In practice, economic characteristics may cluster firms slightly differently from return dynamics, yet the information in $\mathcal{X}$ can still serve as a powerful auxiliary signal to guide clustering in $\bY$. This setting resembles transfer learning: one could apply a debiasing step to explicitly correct the partial misalignment between $\mathcal{X}$-based and $\bY$-based clusters, while still exploiting the shared latent structure. Such an approach would broaden applicability to heterogeneous but related datasets.

The empirical results reveal substantial heterogeneity in factor exposures across groups. Some groups primarily load on size and value factors, while others exhibit strong exposure to profitability or investment. This heterogeneity suggests that enforcing a uniform factor structure across firms may obscure significant cross-sectional variation. The PMTC framework can be extended to accommodate high-dimensional ``factor zoos" by incorporating sparse estimation techniques to select group-specific relevant factors. Sparse estimation techniques such as LASSO can isolate the most relevant factors within each cluster, aligning with the literature on uncommon factors and asset heterogeneity \citep[e.g.,][]{cong2023sparse}. Recognizing that clusters may be driven by distinct factors, the coupled matrix-tensor framework provides a natural setting for factor selection.

\vspace{1cm}
{\onehalfspacing
\bibliographystyle{apalike}
\bibliography{tensorfinance.bib}
}

\clearpage

\appendix
\begin{center}
\large\bfseries Supplementary Material to ``Panel Coupled Matrix-Tensor Clustering Model with Applications to Asset Pricing"
\end{center}
\vspace{1em}
\setcounter{figure}{0} 
\renewcommand{\thefigure}{\thesection\arabic{figure}}

\section{Additional Simulation Studies}

\subsection{Simulation for Tensor Co-clustering} \label{section:appendix_coclustering}

\begin{figure}[h!]
\caption{\footnotesize \textbf{CER for different methods for the tensor block model}}
\vspace{-20pt}
\begin{center}
\adjustbox{center}
{\includegraphics[width=0.8\textwidth]{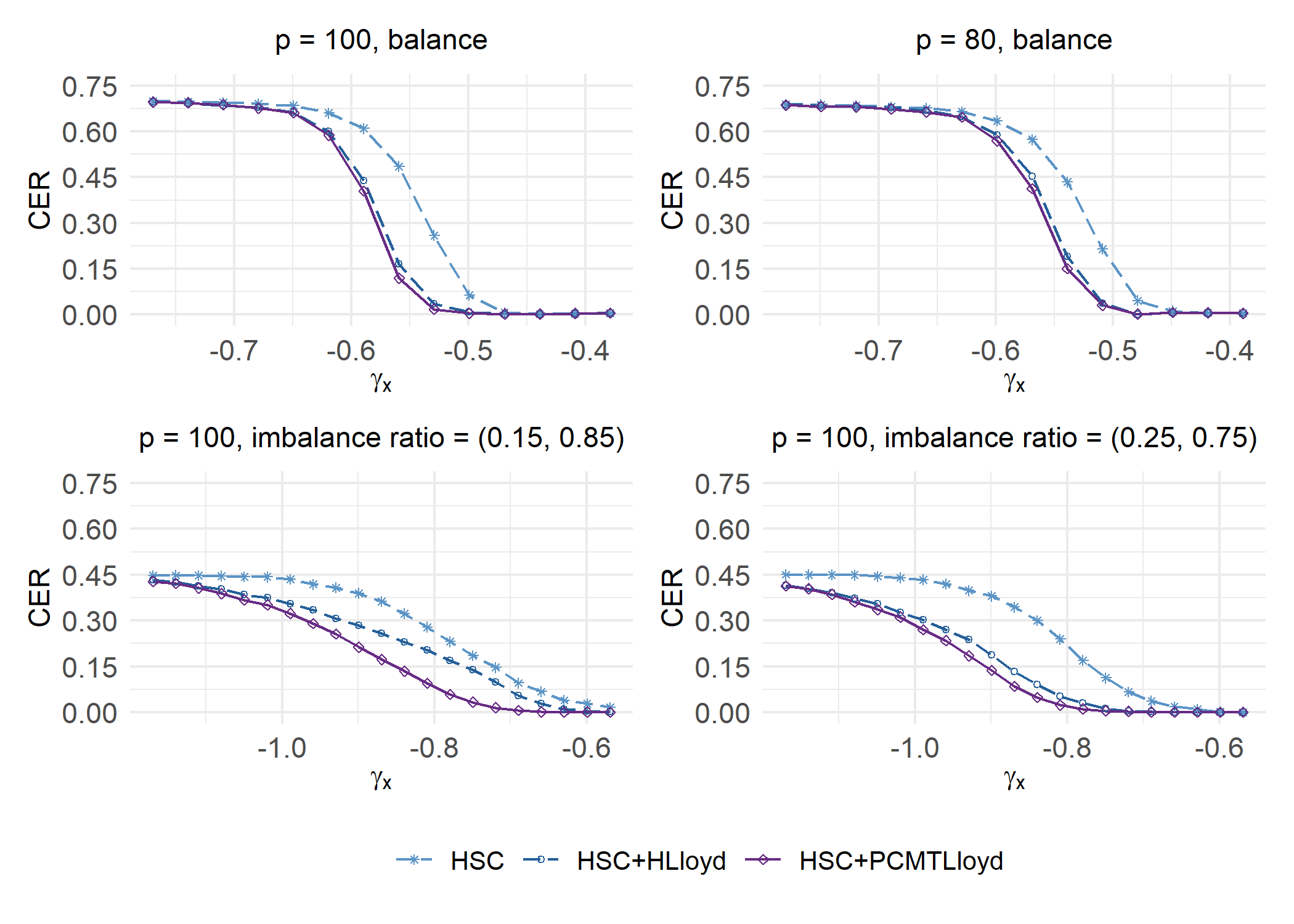}}    
\end{center}
\vspace{-25pt}
\footnotesize{\underline{Note:}
The top two panels display CER for balanced cases; the bottom two panels show CER for imbalanced cases. Both HLloyd \citep{han2022exact} and PMTLloyd use HSC for initialization. Settings: tensor $\cX$ follows the tensor block model \eqref{model:tensorblock}. Results are averaged over 100 repetitions.}
\label{fig-X-coclustering}
\end{figure}

In this subsection, we compare the proposed PMTLloyd algorithm to high-order spectral clustering (HSC) and high-order Lloyd algorithm (HLloyd) \citep{han2022exact} in a tensor co-clustering setting. Consider the Gaussian tensor block model:
\begin{equation}
\label{model:tensorblock}
\cX = \cS \times_{i=1}^{d} M_i + \cE,
\end{equation}
with $\sigma^2 = 1$ and $d = 3$. For balanced clustering, we set $r = 5$ and $p \in \{80, 100\}$. For imbalanced clustering, we set $r = 2$, $p = 100$, with imbalance ratios of $(0.15, 0.85)$ and $(0.25, 0.75)$, meaning each element belongs to group 1 with probability 0.15 (or 0.25) and to group 2 with probability 0.85 (or 0.75). Both PMTLloyd and HLloyd use HSC for initialization.

As anticipated, PMTLloyd uniformly outperforms HLloyd across all settings. The bottom two panels of Figure~\ref{fig-X-coclustering} show that the orthogonal projection in PMTLloyd yields particularly large gains than HLloyd when clusters are imbalanced.

\subsection{Additional Simulation for PCHOOI}

\begin{figure}[h!]
\caption{\footnotesize \textbf{Estimation errors under varying $r$} }
\vspace{-20pt}
\begin{center}
\includegraphics[width=0.75\textwidth]{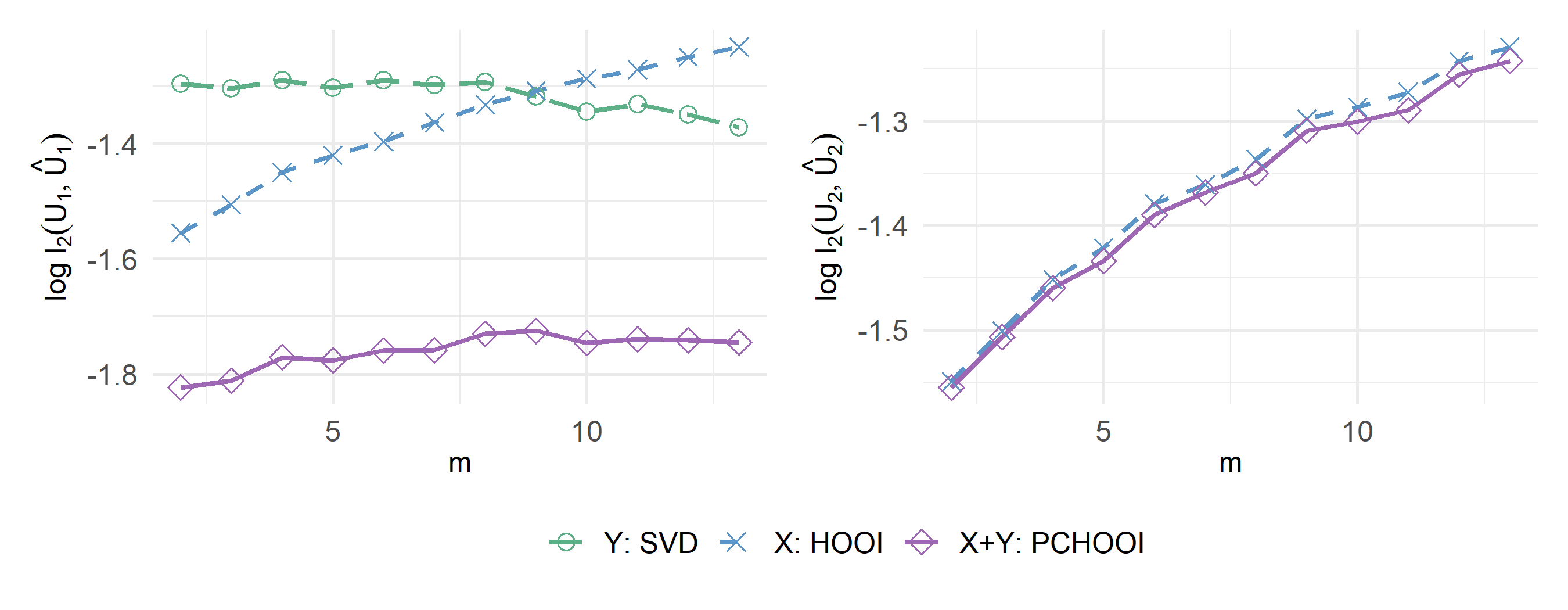}    
\end{center}
\vspace{-15pt}
\footnotesize{\underline{Note:}
The left panel shows errors for $\hbU_1$; the right panel shows errors for $\hbU_2$. Settings: $p_1 = p_2 = 50$, $T = 40$, $\sigma_x = \sigma_y = 1$, $m_1 = m_2 = 5$, $\lambda_{\min}(\mathcal{S}) = c_x \sqrt{p_1 + m_*T}$, and $\lambda_{\min}(\bS_Y) = c_y \sqrt{p_1 + T}$. 
Results, averaged over 100 repetitions, are presented on a log scale.
}
\vspace{-5pt}
\label{PCHOOI_Snorm_r}
\end{figure}

We compare PCHOOI with standard HOOI applied to either $\mathcal{X}$ or $\bY$ alone, using the same data-generating process as in the main text but with $\log C_x = 0$, $\log C_y = 1$, and $r$ varying from 2 to 13. As shown in Figure~\ref{PCHOOI_Snorm_r}, PCHOOI consistently outperforms the alternatives across all values of $r$.

\subsection{Simulation for Clustering and Loading Estimation under Balanced Clusters}\label{section:appendix_balance}

\begin{figure}[h!]
\caption{\footnotesize \textbf{CER for different methods} }
\vspace{-20pt}
\begin{center}
\adjustbox{center}
{\includegraphics[width=0.75\textwidth]{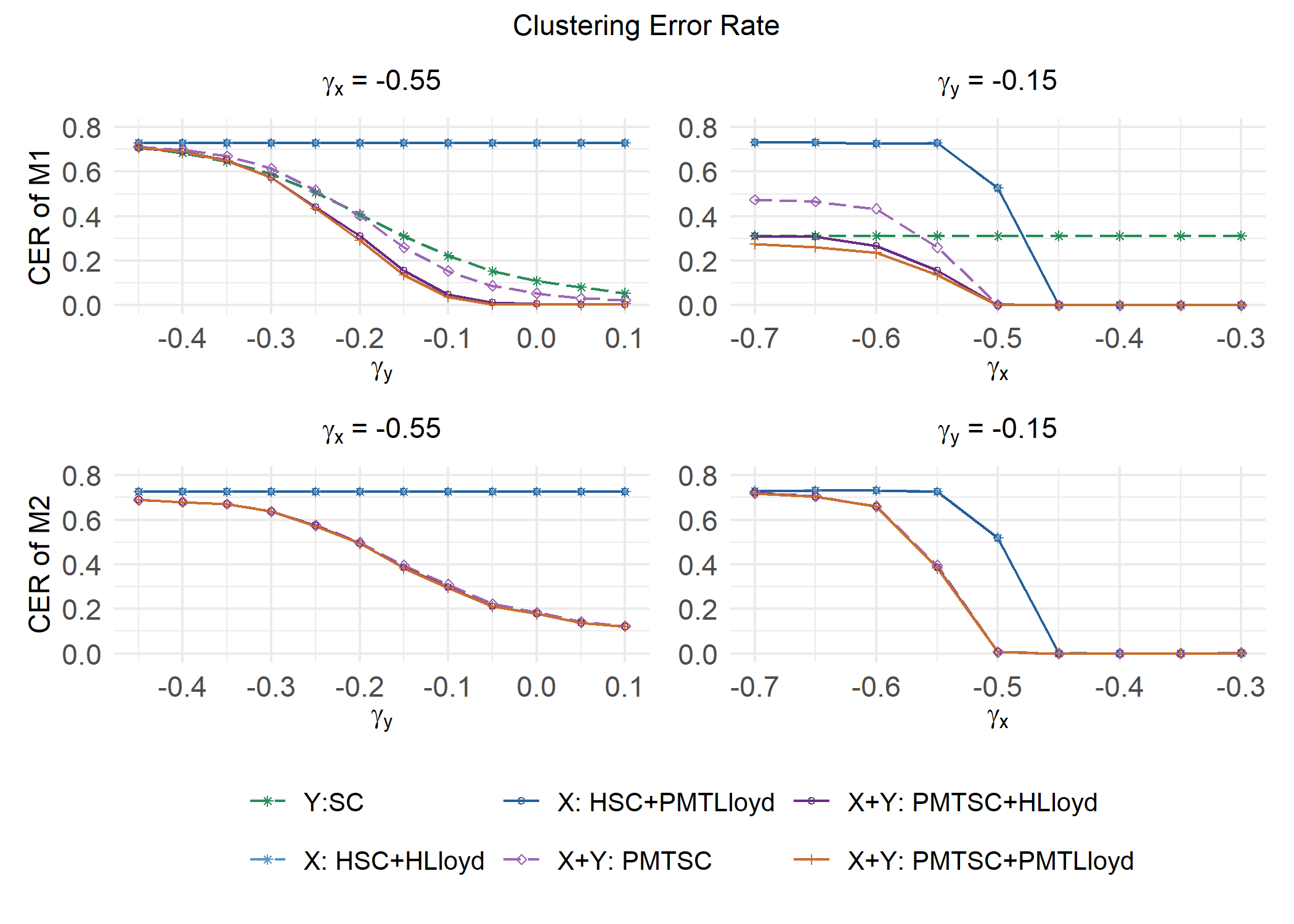}} 
\end{center}
\vspace{-25pt}
\footnotesize{\underline{Note:}
The top two panels display CER for the first mode (coupled); the bottom two panels show CER for the second mode (uncoupled). Settings: $r_1 = r_2 = m_1 = 5$, $p_1 = p_2 = 200$, $T = 120$, with balanced clusters. Results are averaged over 100 repetitions.
}
\label{fig_CER_PMTC_app}
\end{figure}

\begin{figure}[h!]
\caption{\footnotesize 
\textbf{Factor loading estimation error for different methods} }
\vspace{-20pt}
\begin{center}
   \adjustbox{center}
{\includegraphics[width=0.75\textwidth]{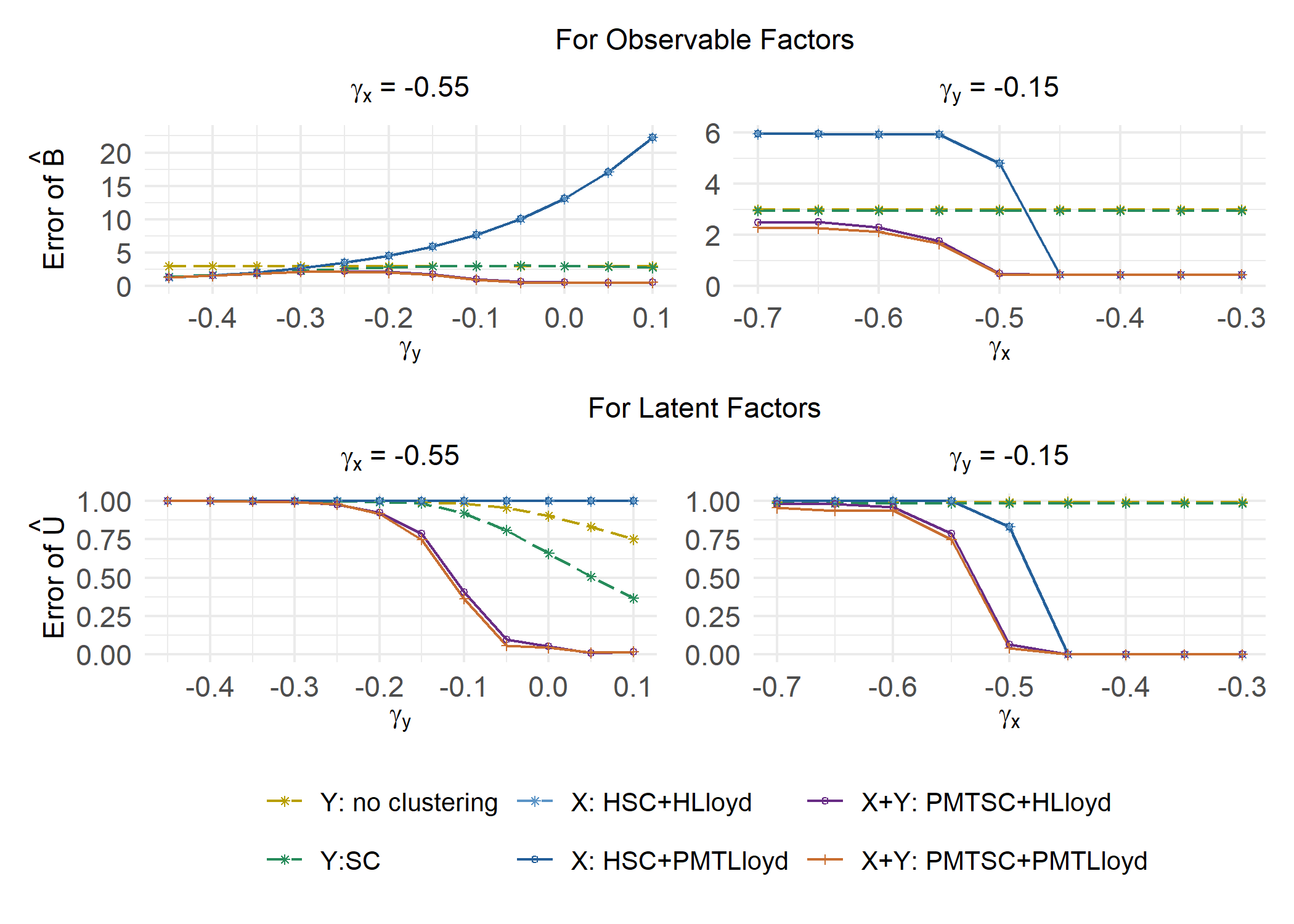}} 
\end{center}
\vspace{-25pt}
\footnotesize{\underline{Note:}
The top two panels display errors for the observed factors case; the bottom two panels show errors for the latent factors case. Results are averaged over 100 repetitions.
}
\vspace{-10pt}
\label{fig-B-err-app}
\end{figure}

Using the same setup as Section~\ref{sec:simulation}, we conduct additional simulations with (i) $\gamma_x = -0.55$ fixed and $\gamma_y$ varying from $-0.45$ to $1$, and (ii) $\gamma_y = -0.20$ fixed and $\gamma_x$ varying from $-0.7$ to $-0.04$. The results are consistent with those in the main text.

\clearpage
\subsection{Simulation for Clustering and Loading Estimation under Imbalanced Clusters}\label{section:appendix_imbalance}

We compare our proposed algorithms to the same benchmarks under imbalanced clustering. Specifically, we set $r_1 = r_2 = 5$ with an imbalance ratio $(0.1, 0.1, 0.15, 0.2, 0.45)$; other parameters remain the same as in the main text. We consider two scenarios: (i) fix $\gamma_x$ and vary $\gamma_y$ from $-0.4$ to $0.1$; (ii) fix $\gamma_y$ and vary $\gamma_x$ from $-0.7$ to $0.4$. The results are consistent with those in the main text.

\begin{figure}[h!]
\caption{\footnotesize \textbf{Clustering Error Rate (CER) under imbalanced clustering}}
\vspace{-20pt}
\begin{center}
\includegraphics[width=0.8\textwidth]{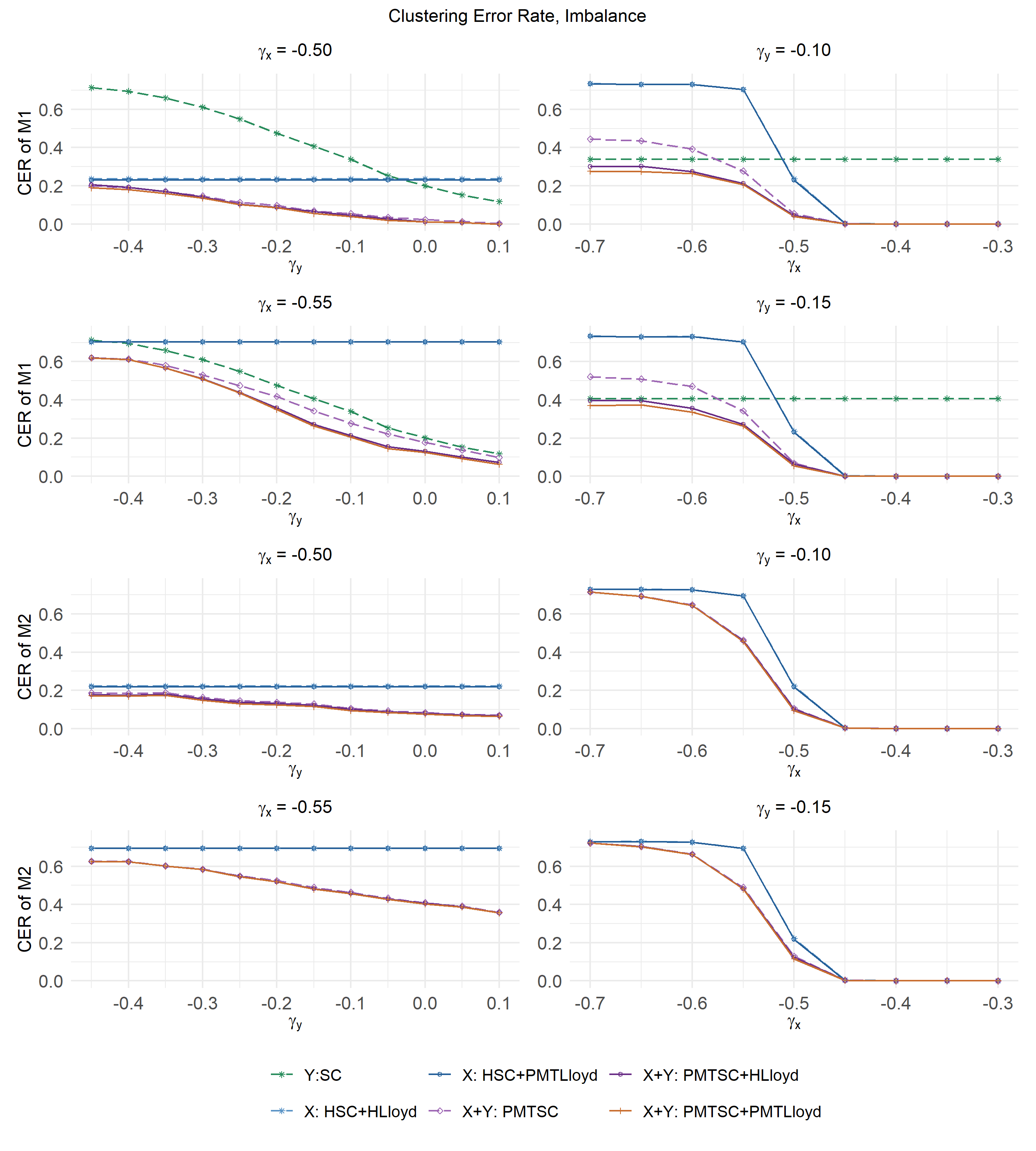} 
\end{center}
\vspace{-15pt}
\footnotesize{\underline{Note:}
The top panels display CER for the first mode (coupled); the bottom panels show CER for the second mode (uncoupled). Settings: $r_1 = r_2 = m_1 = 5$, $p_1 = p_2 = 200$, $T = 120$, with imbalanced clusters. Results are averaged over 100 repetitions.
}
\vspace{-10pt}
\label{fig_CER_PMTC_imbalance}
\end{figure}

\begin{figure}[h!]
\caption{\footnotesize \textbf{Factor loading estimation error under imbalanced clustering} }
\vspace{-20pt}
\begin{center}
\adjustbox{center}
{\includegraphics[width=0.9\textwidth]{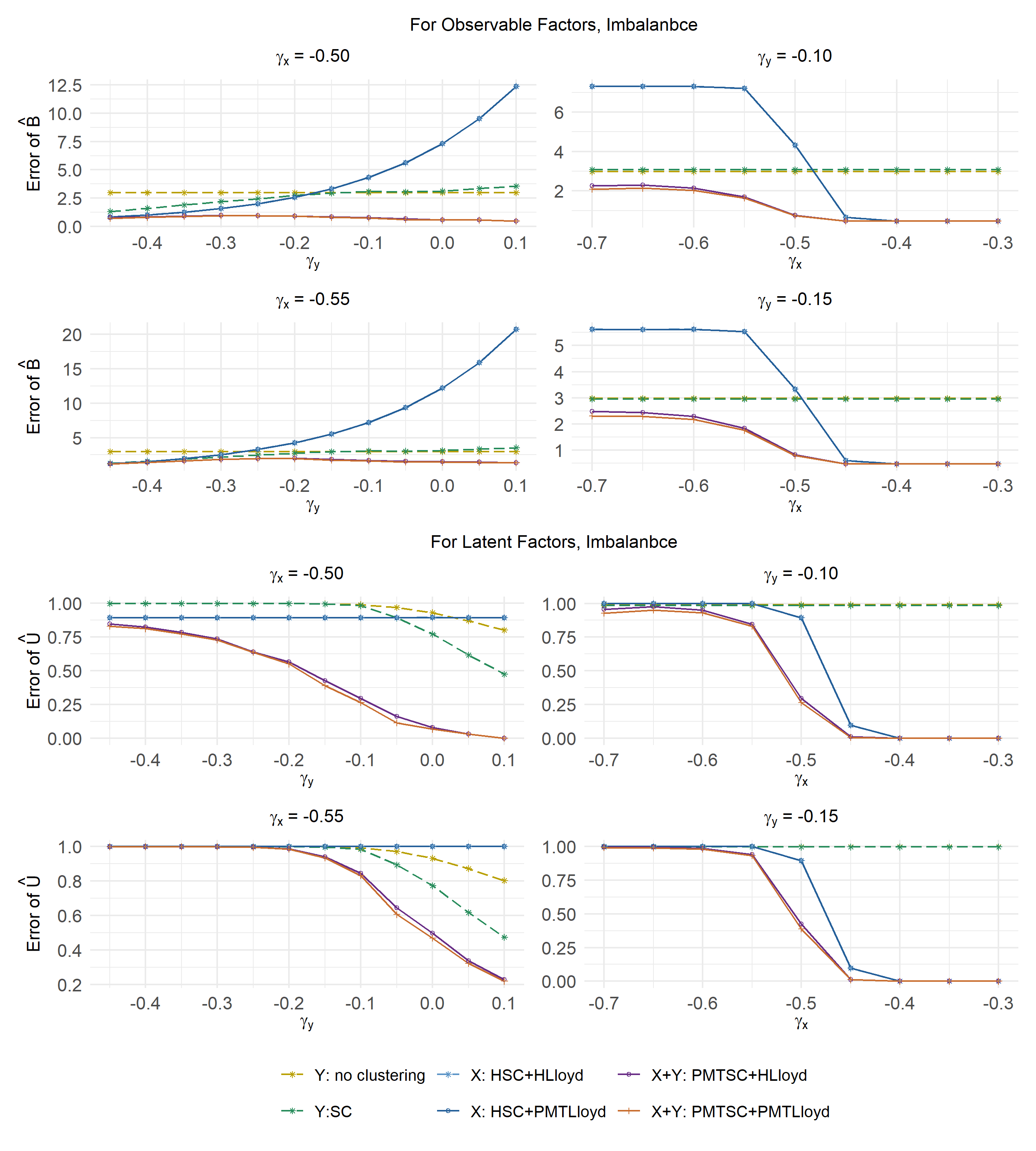}
}    
\end{center}
\vspace{-15pt}
\footnotesize{\underline{Note:}
The top panels display errors for the observed factors case; the bottom panels show errors for the latent factors case. Settings: $ r_1 = r_2 = m_1 = 5$, $p_1 = p_2 = 200$, $T = 120$, with imbalanced clusters. Results are averaged over 100 repetitions.
}
\vspace{-10pt}
\label{fig-B-err-imbalance}
\end{figure}

\clearpage
\subsection{Simulation for Smaller Dimensions} \label{section:appendix_small}

We conduct additional simulations with smaller $p$ and $T$. We set $p = 100$ and $T = 60$, with all other parameters remaining the same as in the main text. The results are consistent with those reported earlier.

\begin{figure}[h!]
\caption{\footnotesize \textbf{CER for smaller dimensions}}
\vspace{-23pt}
\begin{center}
\adjustbox{center}{
\includegraphics[width=0.7\textwidth]{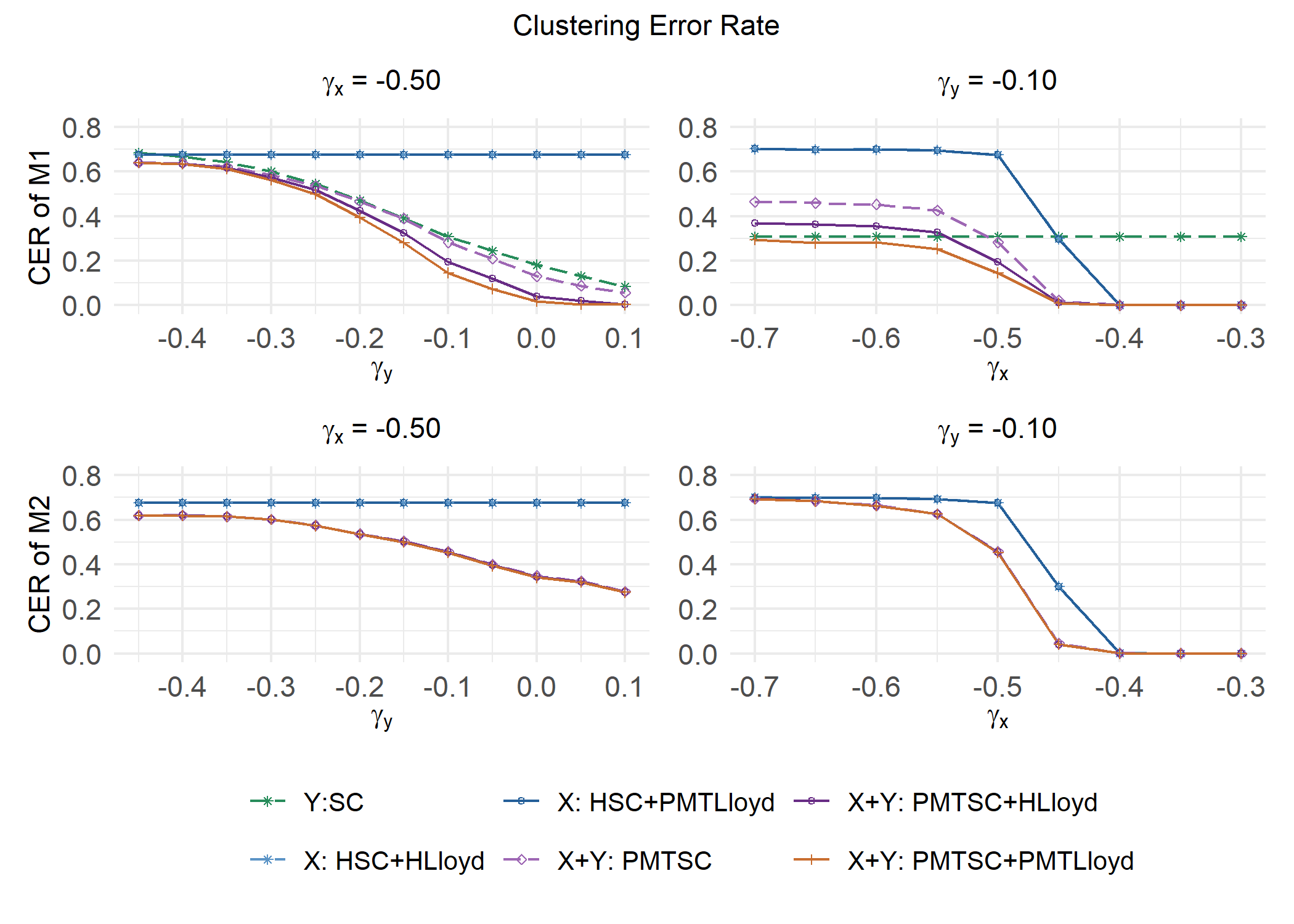}
} 
\end{center}
\vspace{-25pt}
\footnotesize{\underline{Note:}
The top panels display CER for the first mode (coupled); the bottom panels show CER for the second mode (uncoupled). Results are averaged over 100 repetitions.
}
\vspace{-10pt}
\label{fig_CER_PMTC_p100}
\end{figure}

\begin{figure}[h!]
\caption{\footnotesize \textbf{Factor loading estimation error for smaller dimensions}}
\vspace{-23pt}
\begin{center}
\adjustbox{center}
{\includegraphics[width=0.7\textwidth]{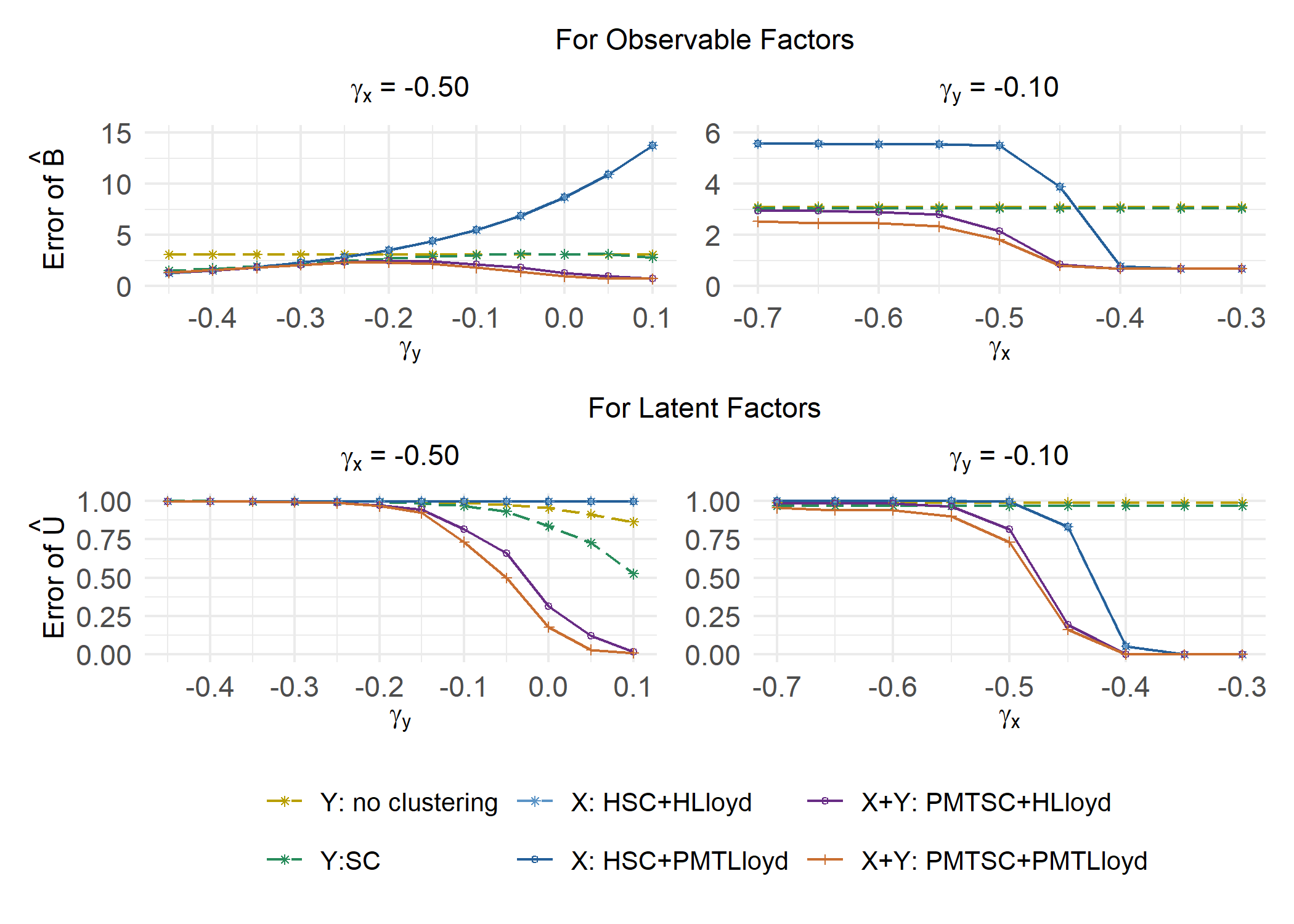}}    
\end{center}
\vspace{-25pt}
\footnotesize{\underline{Note:}
The top panels display errors for the observed factors case; the bottom panels show errors for the latent factors case. Results are averaged over 100 repetitions.}
\vspace{-10pt}
\label{fig-B-err-p100}
\end{figure}

\clearpage
\section{Additional Empirical Application} \label{section:appendix_data}
To rigorously assess the temporal robustness of our proposed method and ensure our results are not artifacts of a specific sample split, we conduct a dynamic stability analysis.
Adopting the rolling window evaluation approach proposed in \cite{zhang2025testing}, we implement a yearly rolling framework where each calendar year serves as the in-sample (INS) estimation period, followed by out-of-sample (OOS) validation in the subsequent year. This procedure iterates across the full horizon, with results averaged over all windows. These complementary schemes offer distinct insights: the 35-year static split captures the long-term stability of latent groups, while the rolling framework tests the model's adaptability to time-varying market dynamics. Jointly, they ensure a robust assessment of our methodology in evolving financial environments.

\begin{table}[h!]
\centering

\label{tab-yearly rebalanced}
\captionsetup[table]{skip=6pt}
\resizebox{\ifdim\width>\linewidth\linewidth\else\width\fi}{!}{
\begin{threeparttable}
\caption*{Table A1: \textbf{Additional Empirical Comparison of Methods}}
\footnotesize
\footnotesize
\begin{tabular}{lcccccccccccc}
\toprule
& \multicolumn{6}{c}{In-Sample (Train) $R^2$ (\%)} & \multicolumn{6}{c}{Out-of-Sample (Validation) $R^2$ (\%)}\\
\cmidrule(lr){2-7}\cmidrule(lr){8-13}
& \multicolumn{3}{c}{Benchmarks} & \multicolumn{3}{c}{\textbf{Our methods}} 
& \multicolumn{3}{c}{Benchmarks} & \multicolumn{3}{c}{\textbf{Our methods}}\\
\cmidrule(lr){2-4}\cmidrule(lr){5-7}\cmidrule(lr){8-10}\cmidrule(lr){11-13}
$r_1$  & Only $\bY$ & BM & ME & $\mathcal{G}_2^{F}$ & PMTSC & PMTLloyd
       & Only $\bY$ & BM & ME & $\mathcal{G}_2^{F}$ & PMTSC & PMTLloyd\\
\midrule
& \multicolumn{12}{c}{yearly rebalanced estimate}\\
\cmidrule(lr){5-11}
1  & 20.1 &      &      &      &      &
   & 14.6 &      &      &      &      & \\
2  & 33.8 &      &      & \textbf{41.8} & 27.6 & 38.1
   & 15.6 &      &      & 23.2 & 19.5 & \textbf{25.4} \\
5  & 41.0 & 27.8 & 32.4 & \textbf{49.8} & 39.3 & 45.3
   & 13.4 & 15.4 & 20.5 & 21.0 & 24.6 & \textbf{25.4} \\
10 & 41.9 & 28.6 & 33.8 & \textbf{52.5} &  42.0 & 48.3
   & 11.8 & 15.0& 20.0 & 23.3 & 15.6 & \textbf{24.2} \\
25 & 43.1 & \multicolumn{2}{c}{42.1} & \textbf{54.6} & 43.2 & 51.2
   & 7.3  & \multicolumn{2}{c}{21.0} & 14.9 &  18.0 & \textbf{22.1} \\
\bottomrule
\end{tabular}
{\footnotesize \underline{Note:}
This table reports the total $R^2$ (\%) for training (In-Sample, INS) and validation (Out-of-Sample, OOS) data across methods. The first three columns (Benchmarks) show baseline approaches: ``Only $Y$" clusters on the returns matrix $\bY$, while ``BM" and ``ME" use univariate sorts with joint characteristic benchmarks. The last three columns (Ours) present our methods. ``$\mathcal{G}_2^{F}$" employs a pre-defined economic grouping for model-2, and ``PMTSC" and ``PMTLloyd" fix the second-mode cluster count at $r_2=6$. We apply a rolling yearly evaluation, averaging OOS results. These methods aim to enhance clustering accuracy by incorporating economic structure.}
\end{threeparttable}
}
\vspace{-10pt}
\end{table}

Results from the rolling framework confirm that PMTLloyd consistently outperforms competing methodologies across both estimation and validation periods. In the standard 10-cluster specification, our approach improves the OOS total $R^2$ by 2.9 percentage points over the returns-only ('Only Y') baseline, underscoring the robust additive value of the characteristics tensor even under time-varying conditions.

\end{document}